\documentclass[aps,prb,twocolumn,showpacs,citeautoscript,superscriptaddress]{revtex4-1}

\usepackage{amsmath}
\usepackage{amssymb}
\usepackage{amsfonts}
\usepackage{graphicx,graphics}

\newcommand{\abs}[1]{\lvert #1 \rvert}
\newcommand{\expect}[1]{\langle #1\rangle}

\newcommand{\bk}{\mathbf{k}}
\newcommand{\bq}{\mathbf{q}}

\begin{document}

\title{Hot-electron cooling by acoustic and optical phonons in monolayers of MoS$_2$
  and other transition-metal dichalcogenides}
\author{Kristen Kaasbjerg}
\email{cosby@fys.ku.dk}
\affiliation{Department of Condensed Matter Physics, Weizmann Institute of
  Science, Rehovot 76100, Israel}
\author{K. S. Bhargavi}
\affiliation{Department of Physics, Karnatak University, Dharwad-580 003,
  Karnataka, India}
\author{S. S. Kubakaddi}
\email{sskubakaddi@gmail.com}
\affiliation{Department of Physics, Karnatak University, Dharwad-580 003,
  Karnataka, India}
\date{\today}

\begin{abstract}
  We study hot-electron cooling by acoustic and optical phonons in monolayer
  MoS$_2$. The cooling power $P$ ($P_e = P/n$) is investigated as a function of
  electron temperature $T_e$ (0--500~K) and carrier density $n$
  ($10^{10}$--$10^{13}$~cm$^{-2}$) taking into account all relevant
  electron-phonon (el-ph) couplings. We find that the cross over from acoustic
  phonon dominated cooling at low $T_e$ to optical phonon dominated cooling at
  higher $T_e$ takes place at $T_e \sim 50-75$~K. The unscreened deformation
  potential (DP) coupling to the TA phonon is shown to dominate $P$ due to
  acoustic phonon scattering over the entire temperature and density range
  considered. The cooling power due to screened DP coupling to the LA phonon and
  screened piezoelectric (PE) coupling to the TA and LA phonons is orders of
  magnitude lower. In the Bloch-Gr{\"u}neisen (BG) regime, $P\sim T_e^4$
  ($T_e^6$) and $P\sim n^{-1/2}$ ($P_e\sim n^{-3/2}$) are predicted for
  unscreened (screened) el-ph interaction. The cooling power due to optical
  phonons is dominated by zero-order DP couplings and the Fr{\"o}hlich
  interaction, and is found to be significantly reduced by the hot-phonon effect
  when the phonon relaxation time due to phonon-phonon scattering is large
  compared to the relaxation time due to el-ph scattering. The $T_e$ and $n$
  dependence of the hot-phonon distribution function is also studied. Our
  results for monolayer MoS$_2$ are compared with those in conventional
  two-dimensional electron gases (2DEGs) as well as monolayer and bilayer
  graphene.
\end{abstract}

\pacs{72.10.-d, 72.80.Jc, 73.63.-b, 81.05.Hd}
\maketitle

\section{Introduction}

Two-dimensional (2D) materials have attracted great interests due to their
interesting physical properties and potential use in next generation
nanoelectronic devices. The most rigorously studied 2D material is graphene
because of its linear energy dispersion relation leading to rich new physics and
zero effective mass of charge carriers with very high room temperature
mobility~\cite{RMP:Graphene,Sarma:RMP}. However, since graphene has zero band
gap it is not well-suited for device applications such as transistors and
detectors. Apart from graphene, monolayers of transition-metal dichalcogenides
(MX$_2$ with M=Mo, W and X=S, Se and Te), atomically thin 2D semiconductors with
a finite band gap, have been recent focus of extensive research
activity~\cite{Strano:NNanoReview,Zhang:NChemReview}. Due to their
semiconducting nature, monolayers of MX$_2$ materials have advantages over
zero-band gap graphene and are suitable for many electronic and photonic
applications. So far, field effect transistors with on/off ratios $> 1\times
10^8$, photo detectors and LEDs based on 2D MX$_2$ materials have been
realized~\cite{Kis:MoS2Transistor,Mueller:SolarEnergy,Herrero:Optoelectronic,Xu:Electrically}.

Monolayer molybdenum disulfide (MoS$_2$) which has a direct band gap of
1.8~eV~\cite{Heinz:ThinMoS2}, is a typical example of these MX$_2$ materials.
Transport properties of monolayer MoS$_2$ are being studied
experimentally~\cite{Kis:MoS2Transistor,Ghosh:Nature,Kim:HighMobility,Balicas:Intrinsic,Kis:Engineering,Eda:TransportProperties,Avouris:Prospects}
and
theoretically~\cite{Kaasbjerg:MoS2,Kaasbjerg:MoS2Acoustic,Dery:SymmetryBased,Fischetti:Mobility,Jena:ChargeScattering,Guinea:Effect},
and most of this work is concentrated on the electron mobility which sets the
upper limit for the operational speed of the electronic devices. Experimentally,
room temperature mobilities in the range 1--200~cm$^2$/Vs in $n$-type monolayer
MoS$_2$ samples have been
reported~\cite{Kis:MoS2Transistor,Hone:Measurement,Kis:Reply}. Dielectric
engineering has been used to achieve the highest mobilities in top gated samples
with high-$\kappa$ gate dielectrics. In this case, the scattering due to
impurities can be drastically suppressed by screening~\cite{Konar:Engineering}
and mobilities close to intrinsic phonon-limited mobility of $\sim 410$~cm$^2$/Vs
can be achieved~\cite{Kaasbjerg:MoS2,Kaasbjerg:MoS2Acoustic}.
\begin{figure}[!b]
  \centering
  \includegraphics[width=0.45\linewidth]{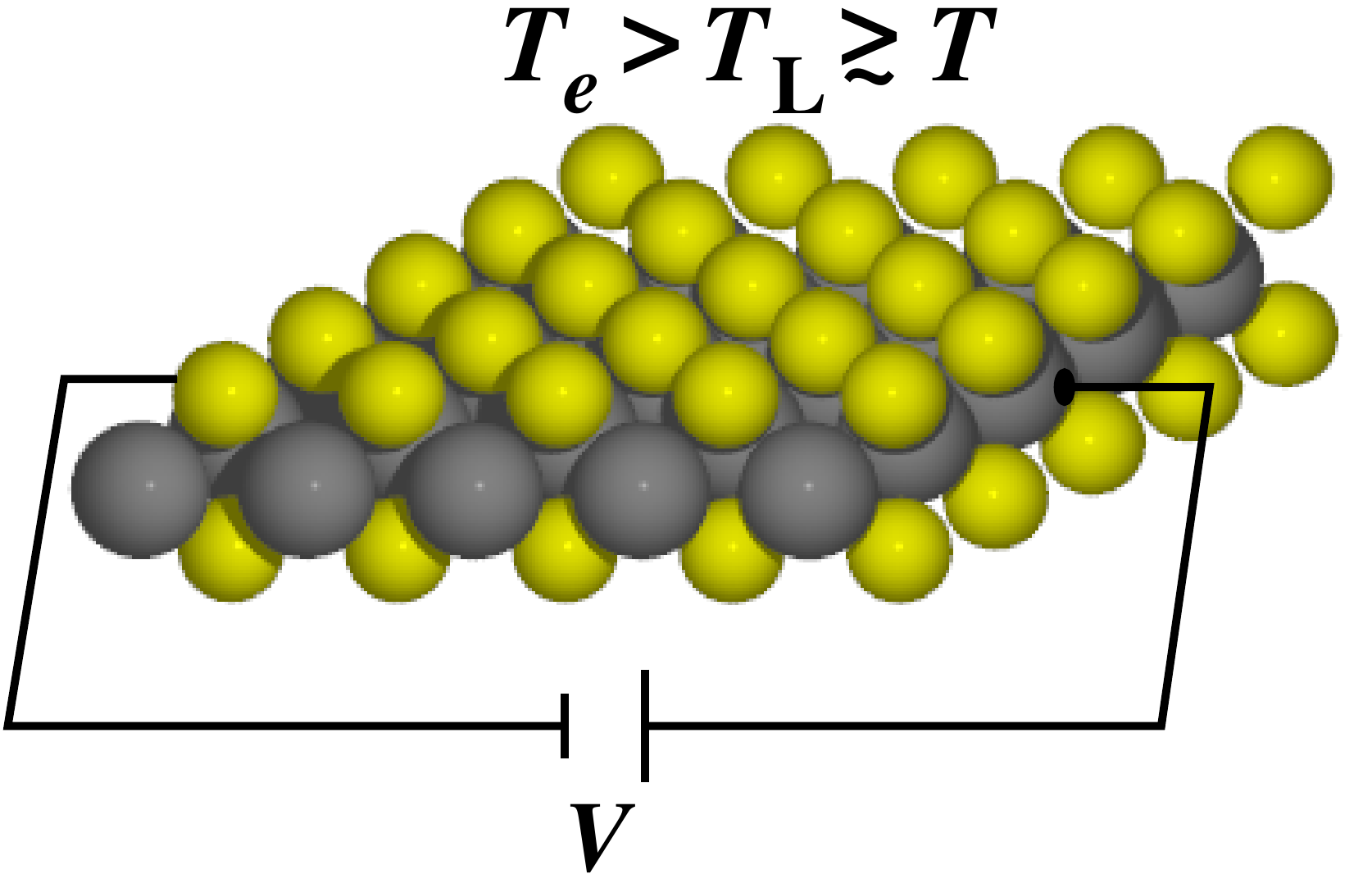}
  \caption{(Color online) Schematic illustration of a biased monolayer MoS$_2$
    transistor. The applied bias $V$ results in a quasi-equilibrated
    hot-electron distribution characterized by a temperature $T_e$ larger than
    the temperature $T_\text{L}$ of the crystal lattice. Due to the power $P=IV$
    dissipated in the device, the lattice may heat up with respect to the
    substrate/device environment held at temperature $T$, i.e. $T_\text{L}>
    T$. This results in a reduction of the cooling power due to the so-called
    hot-phonon effect.}
  \label{fig:mos2}
\end{figure}
Besides, unlike the conventional semiconductor heterostructures, there is no
intrinsic roughness over the 2D plane in atomically thin semiconductors and the
absence of surface roughness, in principle, makes it possible to attain still
higher mobilities. Efforts are still on going to realize the highest possible
room temperature mobilities largely limited by electron-phonon (el-ph) scattering.

In photoexcited samples and samples subject to high electric fields, electrons
are appreciably heated and driven out of equilibrium with the lattice. This is
an important phenomenon as it affects thermal dissipation and heat management
which are key issues in nanoscale electronics and will play a role in any future
MX$_2$ based devices. In addition, hot electrons mediate energy transport which
finds applications in variety of devices such as calorimeters, bolometers,
infrared and THz detectors, and furthermore gives rise to the
photothermoelectric effect observed in monolayer
MoS$_2$~\cite{Gomez:Photothermo}. Understanding the important pathways for
hot-electron cooling is thus of high importance.

An important channel for cooling of hot electrons is by energy transfer to the
host lattice, i.e. phonons. Hot electrons loose their energy by emission of
acoustic phonons at low temperatures and optical phonons at higher temperatures,
and the dependence of the hot-electron relaxation on temperature and carrier
density can provide useful insight into the mechanisms responsible for their
cooling. The study of hot-electron energy relaxation in, e.g., conventional
two-dimensional electron gases (2DEGs) (see, e.g.,
Refs.~\onlinecite{Wiegmann:EnergyLoss,Hess:Electron,Jalabert:Effect,Jalabert:Theory,Ridley:Hot,Harris:EnergyLoss,Butcher:Thermoelectric})
and
monolayer~\cite{Kubakaddi:Interaction,Sarma:EnergyRelaxation,MacDonald:Cooling,Heikkila:ElectronPhonon,Nicholas:Energy,Nicholas:EnergyLoss,Pallecchi:Hot,Finkelstein:Phonon,Schwab:Measurement}
and bilayer~\cite{Heikkila:ElectronPhonon,Kubakaddi:Electronic,Nicholas:Bilayer}
graphene, has been ideal for probing the el-ph coupling since the energy
relaxation, in general, does not depend upon lattice disorder. Recent studies in
graphene, however, have shown that disorder-assisted cooling of hot carriers
through so-called supercollisions plays an important role at higher temperatures
in diffusive
samples~\cite{Levitov:DisorderAssisted,Clerk:ElectronPhonon,McEuen:Photocurrent,Placais:Supercollision,Herrero:Competing}. In
view of these observations, it is important to investigate the different
hot-electron energy relaxation mechanisms in monolayer MoS$_2$.

In this work, we provide a detailed study of the thermal coupling between hot
electrons and the lattice system in monolayer MoS$_2$. To study the cooling of
hot electrons, we take into account all relevant couplings to acoustic and
optical phonons in monolayer
MoS$_2$~\cite{Kaasbjerg:MoS2,Kaasbjerg:MoS2Acoustic}. This includes intravalley
scattering by acoustic phonons via deformation potential and piezoelectric
interaction, intravalley scattering by optical phonons via deformation potential
and the Fr{\"o}hlich interaction, as well as intervalley scattering by both
acoustic and optical phonons via deformation potential interaction. Heating of
the phonons due to relaxation of hot carriers is included through an explicit
solution of the phonon Boltzmann equation. This so-called ``hot-phonon'' effect
is important to account for when electronic reabsorption of excited phonons
becomes a limiting factor for the hot-electron cooling power. Coupling to
surface-polar optical phonons of the substrate/gate dielectric has been
demonstrated to be an important factor for the relaxation of hot electrons in
supported graphene~\cite{Avouris:Cooling,Sarma:Surface}. In monolayer MoS$_2$,
however, the optical phonon energies are significantly lower compared to those
in graphene. The cooling power in MoS$_2$ is therefore more likely dominated by
the intrinsic phonons, why this effect is not considered here. We compare our
results for monolayer MoS$_2$ with those in monolayer and bilayer
graphene~\cite{Kubakaddi:Interaction,Heikkila:ElectronPhonon,Kubakaddi:Electronic}.
As other MX$_2$ monolayers have similar atomic and electronic
structure~\cite{Wirtz:Phonons,Schwing:GiantSO,Yakobson:Quasiparticle}, the
results reported here for MoS$_2$ must be expected to be relevant for other
MX$_2$ variants.

\vspace{.5cm}

\section{Boltzmann theory for the cooling power}
\label{sec:cooling}

Within the framework of Boltzmann transport theory, the evaluation of the
cooling power, in general, requires the solution of coupled electron and phonon
Boltzmann equations for their respective nonequilibrium distribution
functions. However, a major simplification to this problem consists in
assuming that the electrons thermalize among themselves on a fast timescale via
electron-electron scattering. The established quasi equilibrium, maintained by
the applied field, is characterized by a Fermi-Dirac distribution $f_\bk \equiv
f(\varepsilon_\bk) = [e^{(\varepsilon_\bk - \mu) / k_\text{B} T_e } + 1]^{-1}$,
where $\mu$ is the quasi-chemical potential, with a hot-electron temperature
$T_e > T_\text{L}$ larger than the lattice temperature
$T_\text{L}$~\cite{SmithJensen}. Due to the elevated electron temperature, the
energy dissipated to the lattice vibrations may, in addition, drive the phonons
out of equilibrium. The treatment of phonon heating with the phonon Boltzmann
equation is outlined in Sec.~\ref{sec:hot}.

The cooling power $P$ is defined as the rate at which the hot-electron
distribution looses its energy to the phonon system. As this is equivalent to
the rate of change of the energy residing in the phonons, the cooling power (per
sample area $A$) can be obtained as~\cite{SmithJensen}
\begin{equation}
  \label{eq:P}
  P = \frac{1}{A} \sum_{\lambda\bq} \hbar\omega_{\lambda\bq}
      \left( \frac{\partial N_{\lambda\bq}}{\partial t} \right)_\text{coll}^\text{el-ph} ,
\end{equation} 
where $\hbar\omega_{\lambda\bq}$ is the energy of a phonon with branch index
$\lambda$ and 2D wave vector $\bq=(q_x,q_y)$, and $(\partial
N_{\lambda\bq}/\partial t)_\text{coll}^\text{el-ph}$ is the collision integral
which gives the rate of change of the phonon distribution function
$N_{\lambda\bq}$ due to el-ph scattering in the phonon Boltzmann equation (see
Eq.~\eqref{eq:boltzmann} below).

The collision integral due to el-ph scattering is given by
\begin{widetext}
\begin{align}
  \label{eq:collision_elph}
  \left( \frac{\partial N_{\lambda\bq}}{\partial t} \right)_\text{coll}^\text{el-ph}
    & = - \frac{2\pi}{\hbar}
        \sum_{\bk\sigma} \bigg\vert \frac{g_{\bk\bq}^\lambda}{\epsilon(q)} \bigg\vert^2
    \bigg[ 
      f_\bk(T_e) \left\{ 1 - f_{\bk + \bq}(T_e) \right\} N_{\lambda\bq}
    \bigg. \nonumber \\ 
    & \quad 
    \bigg. 
      - f_{\bk + \bq}(T_e) \left\{ 1 - f_\bk (T_e) \right\} \left\{ 1 + N_{\lambda\bq} \right\}
    \bigg]  \delta (\varepsilon_{\bk + \bq} - \varepsilon_\bk - \hbar\omega_{\lambda\bq}) 
\end{align}
\end{widetext} 
where $\bk$ is the electron wave vector, $\sigma$ the electron spin, and the two
terms in the square brackets correspond to processes in which a phonon with
energy $\hbar\omega_{\lambda\bq}$ is absorbed and emitted by the hot-electron
distribution, respectively. For phonons in equilibrium with a phonon bath at
temperature $T$, i.e. $T_\text{L}=T$, the phonon distribution function is given
by the Bose-Einstein distribution, $N_{\lambda\bq}=N_B(T)$.

Inserting the expression for the collision integral in Eq.~\eqref{eq:P}, the
cooling power can be recast in the form~\cite{footnote1}
\begin{align}
  \label{eq:P1}
  P 
    = \sum_\lambda \big[ F_\lambda(T_e) - F_\lambda(T) \big]  ,
\end{align}
where the two terms account for spontaneous emission and stimulated
absorption+emission of phonons, respectively, and the mode-specific function
$F_\lambda$ is defined by
\begin{widetext}
\begin{align}
  \label{eq:F}
  F_\lambda(T) & = \frac{2\pi}{\hbar A}
      \sum_{\bq\bk\sigma} \hbar\omega_{\lambda\bq} 
      \bigg\vert \frac{g_{\bk\bq}^\lambda}{\epsilon(q)} \bigg\vert^2
      N_B(T)
      \left[ f_\bk(T_e) - f_{\bk + \bq}(T_e) \right]
      \delta (\varepsilon_{\bk + \bq} - \varepsilon_\bk - \hbar\omega_{\lambda\bq})  .
\end{align}
\end{widetext}
It follows directly from~\eqref{eq:P1} and~\eqref{eq:F} that in a situation
where the electrons and phonons have equilibrated to a common temperature, $T_e
= T$, the cooling power vanishes as required by detailed balance between the
absorption and emission processes. Furthermore, at $T=0$ where there are no
thermally excited phonons and $N_B(T=0)=0$, the second term in Eq.~\eqref{eq:P1}
vanishes and the cooling power is given entirely by spontaneous emission
processes. At $T\neq 0$, stimulated absorption of phonons will, in general,
dominate stimulated emission processes due to the Fermi factors in
Eq.~\eqref{eq:collision_elph}, resulting in an overall reduction of the cooling
power.

The expression for the cooling power obtained here is completely general and
applies to both acoustic and optical phonons as well as a general electronic
band structure. Furthermore, the expression for the cooling power in
Eqs.~\eqref{eq:P1} and~\eqref{eq:F} holds for a general out-of-equilibrium
phonon distribution function. A rigorous treatment of the hot-phonon effect thus
follows directly with the replacement $N_B(T) \rightarrow N_{\lambda\bq}$ where
$N_{\lambda\bq}$ is the hot-phonon distribution function given below in
Eq.~\eqref{eq:Nhot}.

\subsubsection{General expression for parabolic bands}

For a valley-degenerate 2D semiconductor with parabolic band structure,
$\varepsilon_\bk = \hbar^2k^2/2m$, the $\bk,\bq$ sums can be converted into
integrals, $\sum_{\bk\bq} \rightarrow \tfrac{A^2}{(2\pi)^4} \int \! q \,dq
\,d\theta_\bq \int \!  k \, dk \,d\theta_{\bk\bq}$, and using the $\delta$
function to perform the integration over the polar angle $\theta_{\bk\bq}$
between the two wave vectors, the function $F_\lambda$ in Eq.~\eqref{eq:P1} can
be expressed as
\begin{align}
  \label{eq:F_parabolic}
  F_\lambda(T) & = \frac{g_s g_v A m^{3/2}}{2^{5/2} \pi^3 \hbar^4} 
           \int_0^\infty\!dq \int_0^{2\pi} \! d\theta_\bq 
           \int_{E_0}^\infty \! d\varepsilon_\bk \,
           \bigg\vert \frac{g_{\bk\bq}^\lambda}{\epsilon(q)} \bigg\vert^2
           \nonumber \\
       & \quad \times \frac{\hbar\omega_{\lambda\bq}} 
                           {\sqrt{\varepsilon_\bk - E_0}}
               N_B(T) \left[ f_\bk(T_e) - f_{\bk + \bq}(T_e) \right] ,
\end{align}
where $g_s$ and $g_v$ are the spin and valley degeneracy, respectively, and $E_0
= (\hbar \omega_\bq - \varepsilon_\bq)^2 / 4\varepsilon_\bq$. Due to the
integration over the polar angle $\theta_\bq$, the square of the el-ph coupling
can here be replaced by its angular average $\expect{\abs{g_{\lambda\bq}}^2} =
\tfrac{1}{2\pi} \int \! d\theta_\bq\, \abs{g_{\lambda\bq}}^2$.

\subsection{Carrier energy relaxation rate}

The carrier energy relaxation rate $P(\varepsilon_\bk)$---defined as the net
power flow out of an electronic state---provides information about where in the
hot-electron distribution carriers loose and gain energy via scattering by
phonons. As the rate of change of the hot-electron distribution function is
given by the el-ph collision integral $(\partial f_\bk/\partial
t)_\text{coll}^\text{el-ph}$ from the electron Boltzmann equation, the carrier
energy relaxation rate simply follows by multiplying with the carrier
energy $\varepsilon_\bk$,
\begin{equation}
  \label{eq:P_e}
  P(\varepsilon_\bk) = - \varepsilon_\bk 
    \left( \frac{\partial f_{\bk}}{\partial t} \right)_\text{coll}^\text{el-ph} .
\end{equation}
The full expression for the electron collision integral $(\partial
f_\bk/\partial t)_\text{coll}^\text{el-ph}$ is here omitted and can be found in,
e.g., Ref.~\onlinecite{SmithJensen}. With the above sign convention for
$P(\varepsilon_\bk)$, energy is flowing into the electronic state $\bk$ when
$P(\varepsilon_\bk)<0$, while for $P(\varepsilon_\bk)>0$, energy is flowing out
of the state $\bk$. The carrier energy $\varepsilon^*$ at which
$P(\varepsilon_\bk)$ changes sign, depends on the degeneracy regime of the
electron gas.

It should be noted that when $P(\varepsilon_\bk)$ is summed over $\bk$, the
total cooling power $P= \tfrac{1}{A} \sum_{\bk\sigma} P(\varepsilon_\bk)$ is
obtained. The cooling power obtained in this
way~\cite{Sarma:EnergyRelaxation,Heikkila:ElectronPhonon}, is identical to the
one given in Eqs.~\eqref{eq:P1},~\eqref{eq:F} here.

\section{Hot phonons}
\label{sec:hot}

Heating of phonons due to relaxation of hot carriers becomes important when the
relaxation mechanisms responsible for their equilibration such as, e.g.,
anharmonic phonon-phonon (ph-ph) scattering~\cite{Klemmens:Anharmonic} or
coupling to substrate phonons, are the overall bottleneck for the heat
transport.

For a rigorous treatment of ``hot phonons'' and their impact on the cooling
power, the phonon distribution function must be obtained from the Boltzmann
equation taking into account the nonequilibrium heating due to hot-electron
relaxation as well as the above-mentioned phonon-related damping mechanisms. In
App.~\ref{sec:GF}, we demonstrate the equivalence between the Boltzmann
treatment below and a full quantum-kinetic description within the framework of
the Keldysh nonequilibrium Green function formalism.

In the absence of time-dependent driving terms, the phonon Boltzmann equation
reads~\cite{SmithJensen}
\begin{equation}
  \label{eq:boltzmann}
  \mathbf{v}_{\lambda\bq} \cdot \nabla N_{\lambda\bq} 
  = \left( \frac{\partial N_{\lambda\bq}}{\partial t} \right)_\text{coll}^\text{ph-ph}
    + \left( \frac{\partial N_{\lambda\bq}}{\partial t} \right)_\text{coll}^\text{el-ph},
\end{equation}
where $\mathbf{v}_{\lambda\bq}=\nabla_\bq\omega_{\lambda\bq}$ is the group
velocity of the phonons. Considering a spatial uniform situation, i.e. no
temperature gradients, the left-hand side of the Boltzmann equation is zero,
implying that the two collision terms on the right-hand side must cancel. In
steady state, the rate of increase of the phonon distribution function due to
relaxation of hot electrons is balanced by the decay rate due
ph-ph interactions.

The collision integral due to ph-ph scattering is here described in the
relaxation-time approximation,
\begin{equation}
  \left( \frac{\partial N_{\lambda\bq}}{\partial t} \right)_\text{coll}^\text{ph-ph}
  = - \frac{N_{\lambda\bq} - N_B(T) }{\tau_\text{ph}} ,
\end{equation}
where $\tau_\text{ph}$ is the phonon lifetime due to ph-ph scattering. As a
microscopic treatment of ph-ph interactions is out of the scope of the present
work, we shall here treat the $\tau_\text{ph}$ as a phenomenological parameter.

The collision integral for el-ph scattering in Eq.~\eqref{eq:collision_elph} can
also be written as a relaxation-time expression~\cite{footnote1}
\begin{equation}
  \label{eq:collision_relax}
  \left( \frac{\partial N_{\lambda\bq}}{\partial t}\right)_\text{coll}^\text{el-ph}
  = - \frac{N_{\lambda\bq} - N_B(T_e)}{\tau_{\lambda\bq}} .
\end{equation}
Here, $\tau_{\lambda\bq}$ is the phonon lifetime time due to el-ph scattering
which is given by (see also App.~\ref{sec:GF}) 
\begin{align}
  \label{eq:tau_elph}
  \tau_{\lambda\bq}^{-1} & = \frac{2\pi}{\hbar}
      \sum_{\bk\sigma}
      \bigg\vert \frac{g_{\bk\bq}^\lambda}{\epsilon(q)} \bigg\vert^2
      \left[ f_\bk(T_e) - f_{\bk + \bq}(T_e) \right] \nonumber \\
      & \quad \times \delta (\varepsilon_{\bk + \bq} - \varepsilon_\bk - \hbar\omega_{\lambda\bq}) .
\end{align}

From the relaxation-time expressions for the two collision terms above, it is
evident that the ph-ph and el-ph interactions seek to drive the distribution
function towards Bose-Einstein distributions with temperatures $T$ and $T_e$
of the phonon and electron bath, respectively. This is manifested directly in
the solution to the Boltzmann equation. Solving for the distribution function,
one gets
\begin{equation}
  \label{eq:Nhot}
  N_{\lambda\bq} = 
  \frac{\tau_\text{ph}^{-1} N_B(T) + \tau_{\lambda\bq}^{-1} N_B(T_e)}
       {\tau_\text{ph}^{-1} + \tau_{\lambda\bq}^{-1} } .
\end{equation}
Clearly, the distribution function approaches a Bose-Einstein distribution,
$N_{\lambda\bq} \rightarrow N_B(T)$, characterized by a temperature given by the
substrate/environmental (electron) temperature $T$ ($T_e$) in the limit where
ph-ph (el-ph) scattering dominates the total scattering rate,
$\tau_\text{tot}^{-1} = \tau_\text{ph}^{-1} + \tau_{\lambda\bq}^{-1}$.

A common way to quantify the heating of phonons, is to parametrize the
hot-phonon distribution function in Eq.~\eqref{eq:Nhot} by a Bose-Einstein
distribution,
\begin{equation}
  \label{eq:NhotTeff}
  N_{\lambda\bq} = \frac{1}
      {e^{\hbar\omega_{\lambda\bq}/k_\text{B} T_{\text{eff},\lambda}(q)} - 1} ,
\end{equation}
with the effective phonon temperature $T_{\text{eff},\lambda}(q)$ defined to
yield the \emph{correct} population factor. We here reiterate that the results
for the cooling power given in Sec.~\ref{sec:cooling} hold for a general
out-of-equilibrium phonon distribution function, meaning that the hot-phonon
effect can be taken into account with the replacement $T \rightarrow
T_{\text{eff},\lambda}(q)$.

\section{Electron-phonon interaction}

In extrinsic $n$-type monolayer MoS$_2$, charge carriers reside in the $K,K'$
valleys of the conduction band which are parabolic up to an energy of $\sim
300$~meV and well separated from the satellite valleys inside the Brillouin zone
by a $\sim 300$~meV gap~\cite{Lambrecht:Quasiparticle,Kaasbjerg:MoS2}. At
carrier energies $\varepsilon_k \lesssim 300$~meV, it thus suffices to consider
intra and inter-valley scattering processes in/between the $K,K'$ valleys.

The Hamiltonian for the el-ph interaction in the $K,K'$ valleys takes the
well-known form (with the spin index omitted),
\begin{equation}
  \label{eq:H_elph}
  H_\text{el-ph}  = \sum_{\bk\bq\lambda} g_{\bk\bq}^\lambda
     c_{\bk+\bq}^\dagger c_{\bk}^{\phantom\dagger} (a_{\bq\lambda}^\dagger +
     a_{-\bq\lambda})  ,
\end{equation}
where $g_{\bk\bq}^\lambda$ is the el-ph coupling between the Bloch states with
wave vector $\bk$ and $\bk + \bq$.

In the following we assume that the coupling constant is independent on $\bk$
and write it in the general form
\begin{equation}
  \label{eq:g}
  g_{\lambda\bq} = \sqrt{\frac{\hbar}{2A\rho\omega_{\lambda\bq}}}
                    M_{\lambda\bq} ,
\end{equation}
where $A$ is the area of the monolayer, $\rho$ is the mass density, and
$\omega_{\lambda\bq}$ the phonon dispersion. The coupling matrix element
$M_{\lambda\bq}$ depends on the phonon branch index $\lambda$ as well as the
coupling mechanism.

A detailed analysis of the el-ph couplings in the $K,K'$ valleys of the
conduction band in monolayer MoS$_2$ has been given by some of us in
Refs.~\onlinecite{Kaasbjerg:MoS2,Kaasbjerg:MoS2Acoustic}. For completeness, we
here briefly summarize the couplings to the intra and inter-valley acoustic and
optical phonons.

\subsection{Acoustic phonons}

Due to the lack of inversion symmetry in the hexagonal lattice of monolayer
MoS$_2$, the coupling to the in-plane transverse (TA) and longitudinal (LA)
acoustic phonons with linear dispersion $\omega_{\lambda\bq}=c_\lambda q$
and sound velocity $c_\lambda$, has contributions from both the deformation
potential (DP) and the piezoelectric (PE) coupling mechanisms, 
\begin{equation}
  \label{eq:M_ac}
  M_{\lambda\bq} = M_{\lambda\bq}^\text{DP} + M_{\lambda\bq}^\text{PE} .
\end{equation}
The simultaneous coupling via the two mechanisms gives rise to interference
between them when they are in phase implying that $\abs{M_{\lambda\bq}}^2 \neq
\abs{M_{\lambda\bq}^\text{DP}}^2 + \abs{M_{\lambda\bq}^\text{PE}}^2$. On the
contrary, when the two coupling mechanisms are out of phase, i.e. one is real
and the other complex, they do not interfere, $\abs{M_{\lambda\bq}}^2 =
\abs{M_{\lambda\bq}^\text{DP}}^2 + \abs{M_{\lambda\bq}^\text{PE}}^2$, and can be
treated as separate couplings. For monolayer MoS$_2$, the DP and PE interactions
are in (out of) phase for the TA (LA) mode in the long-wavelength
limit~\cite{Kaasbjerg:MoS2Acoustic}.

For the deformation potential coupling, the matrix element is given by
\begin{equation}
  \label{eq:M_DP}
  M_{\lambda\bq}^{\text{DP}} = \Xi_\lambda q ,
\end{equation}
where $\Xi_\lambda$ is the effective deformation potential. It has been shown
that in the long-wavelength limit, the deformation potential interaction for
the TA and LA phonons is completely dominated by umklapp and normal processes,
respectively~\cite{Kaasbjerg:MoS2Acoustic}.

For the piezoelectric interaction, the matrix element is given
by~\cite{Kaasbjerg:MoS2Acoustic}
\begin{equation}
  \label{eq:M_PE}
   M_{\lambda\bq}^{\text{PE}} = \frac{e_{11} e}{\epsilon_0}
   q \times \text{erfc}(q \sigma / 2) A_\lambda(\hat{\bq}) ,
\end{equation}
where $e_{11}$ is the piezoelectric constant, $\epsilon_0$ is the vacuum
permittivity, erfc is the complementary error function, $\sigma$ is the
effective width of electron wave function, and $A_\lambda(\hat{\bq})$ is an
anisotropy factor accounting for the directional dependence of the piezoelectric
interaction. It is given, respectively, for the TA and LA phonons by
$A_\text{TA}(\hat{\bq}) = -\sin 3\theta_\bq$ and $A_\text{LA}(\hat{\bq})= \cos
3\theta_\bq$, where $\theta_\bq$ is the polar angle of $\bq$ with respect to the
lattice orientation, and the angular average of its absolute square is
$\expect{A_\lambda^2}=1/2$.

It is worth noticing that contrary to the situation in 3D bulk system where
$M_{\lambda\bq}^\text{PE}\sim\text{const.}$ in the long-wavelength
limit~\cite{Mahan}, the matrix element for the piezoelectric interaction in a 2D
lattice goes as $M_{\lambda\bq}^\text{PE} \sim q$ ($\text{erfc}(q\sigma/2)
\approx 1$ for $q\rightarrow 0$). In a 2D material, the deformation potential
and piezoelectric interactions thus have the same $q$ dependence in the
long-wavelength limit.

\subsection{Optical phonons}

For optical-phonon scattering, both zero and first-order deformation potential
interaction with the respective matrix elements given by
\begin{equation}
  \label{eq:M_optical}
  M_{\lambda\bq} = D_\lambda^0  \quad \text{and} \quad 
  M_{\lambda\bq} = D_\lambda^1 q ,
\end{equation}
are considered. In monolayer MoS$_2$, intra and intervalley phonons couple via
both types~\cite{Kaasbjerg:MoS2}. For optical phonons where the lattice
vibration results in a relative atomic displacement inside the unit cell, the
short-range potential giving rise to the deformation-potential interaction is to
a large extent dominated by umklapp processes.

The interaction with the polar LO phonon which originates from the macroscopic
electric field set up by its lattice vibration is described by the Fr{\"o}hlich
interaction~\cite{Madelung}. In 2D materials, the Fr{\"o}hlich interaction is
given by~\cite{Kaasbjerg:MoS2}
\begin{align}
  \label{eq:frohlich}
  g_\text{Fr}(q) & = \sqrt{\frac{e^2 W \hbar\omega_\text{LO}}{2\epsilon_0 A }}
                   \left(
                     \frac{1}{\varepsilon_\infty} - 
                     \frac{1}{\varepsilon_0}
                   \right)^{1/2} \text{erfc}(q \sigma / 2) \nonumber \\
      &  = \frac{g_\text{Fr}}{\sqrt{A}} \text{erfc}(q \sigma / 2) ,
\end{align}
where $W$ is the atomic thickness of the monolayer, and $\varepsilon_\infty$ and
$\varepsilon_0$ are the high-frequency optical and static dielectric constants,
respectively. Instead of evaluating the interaction from the dielectric
constants which are not well established for monolayer MoS$_2$, we here use the
value for the coupling constant $g_\text{Fr}$ obtained in
Ref.~\onlinecite{Kaasbjerg:MoS2}.

\subsection{Screening of the el-ph interactions}
\label{sec:screening}

The effect of screening on the el-ph interaction has recently been discussed by
some of us in Ref.~\onlinecite{Kaasbjerg:MoS2Acoustic}. There, it was shown that
screening of normal and umklapp processes is qualitatively different, with the
screening strength at short wavelengths, i.e. umklapp processes, being strongly
reduced compared to long-wavelength screening.

The contribution to the el-ph interaction from normal and umklapp processes
depends on both the phonon mode and the coupling mechanism implying that the
el-ph couplings are affected differently by carrier screening. For example, the
deformation potential interactions with the long-wavelength TA and LA phonons
are dominated by umklapp and normal processes, respectively, whereas that for
the optical phonons is dominated by umklapp processes only. On the other hand,
the long-range piezoelectric and Fr{\"o}hlich interactions which arise from a
macroscopic polarization of the crystal lattice~\cite{Mahan}, are purely
long-wavelength coupling mechanisms and hence dominated by normal processes.

As free-carrier screening of umklapp processes is
weak~\cite{Kaasbjerg:MoS2Acoustic}, we shall here leave el-ph couplings
dominated by umklapp processes unscreened.

For the screening of the long-wavelength components of the acoustic el-ph
interaction, we consider two sources of screening; i) static screening due to
the 2D carrier density $n$, and ii) background screening from the dielectric
surroundings (substrate, gate dielectrics etc). Dynamical screening of the
Fr{\"o}hlich interaction is weak~\cite{Maldague:ManyBody} due to the large
frequency of the LO phonon and is here neglected.

With the static screening of the 2DEG described at the level of
finite-temperature RPA theory, the total dielectric function can be expressed
as~\cite{Stern:2D},
\begin{equation}
  \label{eq:epsilon}
  \epsilon(q, T, \mu) = \kappa - \frac{e^2}{2\epsilon_0q} \chi^0(q,T, \mu) ,
\end{equation}
where $\kappa$ is an effective dielectric constant of the surroundings,
$\chi^0(q, T,\mu)$ is the finite-temperature polarizability of the 2DEG. The
polarizability is obtained following the approach of
Maldague~\cite{Maldague:ManyBody},
\begin{equation}
  \label{eq:maldague}
  \chi^0(q, T,\mu) = \int_0^\infty \! d\mu' \, 
      \frac{\chi^0(q, 0, \mu')}{4k_\text{B}T \cosh^2{\frac{\mu -
            \mu'}{2k_\text{B}T}}}  ,
\end{equation}
where $\chi^0(q, 0, \mu)$ is the zero-temperature RPA
polarizability~\cite{Stern:2D}. The finite-temperature polarizability is
evaluated numerically following Ref.~\onlinecite{Flensberg:Plasmon}.

Which of the two screening mechanisms that dominates the dielectric function in
Eq.~\eqref{eq:epsilon} depends on the screening strength of the 2DEG. For a
degenerate 2DEG, the dielectric function can be written $\epsilon(q) = \kappa +
q_\text{TF}/ q$, where $q_\text{TF} = g_s g_v m e^2 / 4\pi \epsilon_0 \hbar^2$
is the Thomas-Fermi wave vector. In the strong screening limit, $q_\text{TF} /
k_F \gg \kappa$ implying that $\epsilon \approx q_\text{TF} / q$, i.e. screening
is governed by the 2DEG. In the high-temperature nondegenerate regime, the
screening wave vector is given by the Debye-H{\"u}ckel wave vector $q_\text{D} =
ne^2 / 2\epsilon_0 k_\text{B}T$, and background screening will typically
dominate, $\epsilon \approx \kappa$.
\begin{table}[!t]
\begin{ruledtabular}
\begin{tabular}{lcc}
Parameter &  Symbol  & Value  \\ 
\hline                                     
Lattice constant               &   $a$                     &   3.14 \AA            \\
Ion mass density               &   $\rho$                  &   $3.1\times 10^{-7}$ g/cm$^2$ \\
Effective electron mass        &   $m^*$                   &   0.48 $m_e$          \\
Valley degeneracy              &   $g_v$                   &   2           \\
Effective layer thickness      &   $\sigma$                &   5.41 \AA            \\
Piezoelectric constant         &   $e_{11}$                &   $3.0 \times 10^{-11}$~C/m \\
Transverse sound velocity      &   $c_\text{TA}$           &   $4.2 \times 10^3$ m/s  \\
Longitudinal sound velocity    &   $c_\text{LA}$           &   $6.7 \times 10^3$ m/s  \\
Acoustic deformation potentials\\
TA                             &   $\Xi_\text{TA}$         &   $1.5$ eV       \\
LA                             &   $\Xi_\text{LA}$         &   $2.4$ eV       \\
TA                             &   $D_{\mathbf{K},\text{TA}}^1$       &   $5.9$ eV       \\
LA                             &   $D_{\mathbf{K},\text{LA}}^1$       &   $3.9$ eV       \\
Optical deformation potentials\\ 
TO                             &  $D_{\mathbf{\Gamma},\text{TO}}^1$   &   $4.0$ eV       \\
TO                             &  $D_{\mathbf{K},\text{TO}}^1$        &   $1.9$ eV       \\
LO                             &  $D_{\mathbf{K},\text{LO}}^0$        &   $2.6 \times 10^8$ eV/cm  \\
Homopolar                      &  $D_{\mathbf{\Gamma},\text{HP}}^0$   &   $4.1 \times 10^8$ eV/cm  \\
Fr{\"o}hlich interaction       \\
LO                             &  $g_\text{Fr}$            &   $286$ meV Ang   \\  
Phonon energies                \\ 
TA                             &  $\hbar\omega_{\mathbf{K},\text{TA}}$       & 23 meV \\
LA                             &  $\hbar\omega_{\mathbf{K},\text{LA}}$       & 29 meV \\
TO                             &  $\hbar\omega_{\mathbf{\Gamma},\text{TO}}$  & 48 meV    \\
                               &  $\hbar\omega_{\mathbf{K},\text{TO}}$       & 47 meV    \\
LO                             &  $\hbar\omega_{\mathbf{\Gamma},\text{LO}}$  & 48 meV    \\
                               &  $\hbar\omega_{\mathbf{K},\text{LO}}$       & 41 meV    \\
Homopolar                      &  $\hbar\omega_\text{HP}$  &   50 meV\\
\end{tabular}
\end{ruledtabular}
\caption{Material parameters for single-layer MoS$_2$ adopted from 
  Refs.~\onlinecite{Kaasbjerg:MoS2,Kaasbjerg:MoS2Acoustic}. The
  $\mathbf{\Gamma}/\mathbf{K}$-subscripts indicate intra/intervalley phonons.}
\label{tab:parameters}
\end{table}

\section{Results}

In the following the cooling power in $n$-type monolayer MoS$_2$ is studied
using the material parameters listed in Tab.~\ref{tab:parameters}. We have
evaluated the cooling power numerically as a function of hot-electron
temperature $T_e$ and carrier density $n$ at temperatures $T_e < 500$~K and
densities $10^{10}$--$10^{13}$~cm$^{-2}$ and supplement by analytic
considerations for the limiting behavior at low temperatures where the cooling
power is dominated by acoustic phonons. Dielectric background screening is only
included where mentioned explicitly, otherwise $\kappa=1$. It should be
mentioned that in most of the figures below we show the cooling power per
electron defined by
\begin{equation}
  \label{eq:Pe}
  P_e = P / n ,
\end{equation}
instead of the cooling power per sample area $P$ defined in Eq.~\eqref{eq:P}.

The results presented in the following have been obtained under the following
assumptions for the phonon relaxation due to ph-ph scattering. The acoustic
phonons are assumed to equilibrate with substrate phonons on a fast time scale
such that $\tau_\text{ph} \ll \tau_{\lambda\bq}$, implying that they remain in
thermal equilibrium with the environmental substrate phonons with distribution
function $N_\text{ac}=N_B(T)$. On the other hand, equilibration of optical
phonons is assumed to take place on a slower time scale governed by anharmonic
ph-ph scattering which allows the phonons to be driven out of equilibrium. In
order account for the effect of hot phonons on the cooling power, the
distribution function for the optical phonons $N_\text{op}=N_B(T_\text{eff})$ is
obtained as outlined in Sec.~\ref{sec:hot}.

\subsection{Cooling by acoustic phonons at low $T_e$}

At low temperatures where the thermal energy of the electron distribution is
much smaller than the optical phonon energies, $k_\text{B}T_e \ll
\hbar\omega_\lambda$, the cooling power is dominated by acoustic-phonon
scattering. In this regime, the cooling power can be described by the generic
power-law behavior~\cite{Heikkila:ElectronPhonon}
\begin{equation}
  \label{eq:P_acoustic}
  P = \Sigma(\mu,T_e) (T_e^\delta - T^\delta)
\end{equation}
where $\Sigma$ is an \emph{effective} coupling constant for all the acoustic
el-ph couplings that depends on the chemical potential $\mu$ and the electron
temperature, and $\delta$ is the exponent of the power law which overall
decreases with increasing temperature. These dependencies of $\Sigma$ and
$\delta$ are determined by the function $F_\lambda$ in Eq.~\eqref{eq:F}.

The power-law behavior for the cooling power due to acoustic phonons is
characterized by a crossover between two cooling regimes at $T_e \sim
T_\text{BG}$, where $T_\text{BG}$ is the Bloch-Gr{\"u}neisen (BG) temperature
defined as the temperature at which the thermal energy equals the phonon energy
for full backscattering at the Fermi surface, i.e. $k_\text{B}T_\text{BG} =
2\hbar c_\lambda k_F$ where $k_F$ is the Fermi wave vector. In monolayer
MoS$_2$, the BG temperature for the TA (LA) phonon is $T_\text{BG}\approx
11\sqrt{\tilde{n}} \; \mathrm{K}$ ($\approx 18\sqrt{\tilde{n}} \; \mathrm{K}$)
with the density $\tilde{n}$ in units of $10^{12}$~cm$^{-2}$, thus significantly
lower than the BG temperatures in mono- and bilayer
graphene~\cite{Kaasbjerg:Unraveling}.

In the BG regime $T_e < T_\text{BG}$, the thermal smearing of the electronic
distribution function is smaller than the phonon energy for backscattering at
the Fermi surface. This leads to Pauli blocking of emission processes with wave
vectors $q \sim 2 k_F$ implying that phonon emission is restricted to
small-angle scattering with low phonon energies. As a consequence, the cooling
power increases with a larger value of $\delta$ in the BG regime as compared to
the high-temperature equipartition (EP) regime $T_e > T_\text{BG}$ ($\delta \sim
1$) where the phase space for emission processes is not restricted by Pauli
blocking.

\subsubsection{Analytic low-temperature limits}

We start by obtaining analytic limits for the cooling power due to the different
coupling mechanisms in the low-temperature BG regime. 

In the extreme BG limit, $T_e \ll T_\text{BG}$, the phonon wave vector is
restricted to values $q \ll 2k_F$. Together with the condition $T_F \gg
T_\text{BG}$ where $T_F$ is the Fermi temperature (which is equivalent to $v_F
\gg c_\lambda$), this implies $\hbar\omega_{\lambda \bq} \ll E_F$ and we can
approximate as $f(\varepsilon_\bk) - f(\varepsilon_\bk +
\hbar\omega_{\lambda\bq}) \approx \hbar\omega_{\lambda\bq}
\delta(\varepsilon_\bk - E_F)$, $E_0 \rightarrow 0$ and $\epsilon(q) \approx
q_\text{TF}/ q$. With these approximations inserted in
Eq.~\eqref{eq:F_parabolic}, we find for the low-temperature limits due to
\emph{unscreened} deformation potential interaction
\begin{align}
  \label{eq:Sigma_TADP}
  \Sigma_\lambda^\text{DP} & = \frac{g_s g_v \pi^2 \Xi_\lambda^2 m^{3/2} k_\text{B}^4} 
      {2^{1/2} 60\hbar^5 \rho c_\lambda^3 E_F^{1/2}} \sim n^{-1/2} \quad \text{and}
      \quad \delta = 4,
\end{align}
\emph{screened} deformation potential interaction
\begin{align}
  \label{eq:Sigma_LADP}
  \Sigma_\lambda^\text{DP} & = \frac{2 g_s g_v \pi^4 \Xi_\lambda^2 m^{3/2} k_\text{B}^4} 
      {2^{1/2} 63 \hbar^5 \rho c_\lambda^3 E_F^{1/2}}  S_\lambda \sim n^{-1/2}
      \quad \text{and}
      \quad \delta = 6 ,
\end{align}
and \emph{screened} piezoelectric interaction
\begin{align}
  \label{eq:Sigma_PZ}
  \Sigma_\lambda^\text{PE} & = \frac{g_s g_v \pi^4 (e e_{11} / \epsilon_0)^2 m^{3/2} k_\text{B}^4}
       {2^{1/2} 63 \rho\hbar^5 c_\lambda^3 E_F^{1/2}} S_\lambda \sim n^{-1/2} 
       \nonumber \\ & \quad \text{and} \quad \delta = 6 ,
\end{align}
respectively, where the constant
\begin{align}
  S_\lambda & = \left(
                \frac{4\pi\epsilon_0\hbar k_\text{B}}{g_s g_ve^2 m c_\lambda} 
                \right)^2 
\end{align}
originates from the screening of the el-ph interaction and therefore does not
appear in Eq.~\eqref{eq:Sigma_TADP} for unscreened deformation potential
interaction.

From the above results, we have that $P \sim T_e^4$ for scattering via
unscreened deformation potential coupling. On the other hand, coupling to
acoustic phonons via screened deformation potential and screened piezoelectric
coupling gives $P \sim T_e^6$. As the 2DEG screening function is independent of
the density, $P \sim n^{-1/2}$ ($P_e\sim n^{-3/2}$) for both the unscreened and
screened interactions. The temperature and density dependence obtained here may
be compared with those in conventional 2DEG systems as well as monolayer and
bilayer graphene.

In conventional 2DEGs where phonons are considered to be 3D, the temperature
dependence is $P \sim T_e^5$ ($P \sim T_e^7$) for unscreened (screened)
deformation potential coupling~\cite{Harris:EnergyLoss,Butcher:Thermoelectric}
and $P \sim T_e^3$ ($P \sim T_e^5$) for unscreened (screened) piezoelectric
scattering~\cite{Harris:EnergyLoss}. The difference in the power-law behavior
between monolayer MoS$_2$ and conventional 2DEGs with respect to the deformation
potential coupling can be attributed to the 2D nature of the phonons in the
former. However, the difference in the power law for $P$ due to piezoelectric
coupling can be attributed not only to the reduced dimensionality of the
phonons, but also the different $q$ dependencies of the matrix elements (see
Eq.~\eqref{eq:M_PE} and the discussion following it).

In monolayer and bilayer graphene where phonons are
2D~\cite{Kubakaddi:Interaction,Heikkila:ElectronPhonon,Kubakaddi:Electronic},
the power law for unscreened deformation potential coupling is $P\sim T_e^4$
which is the same as our result for monolayer MoS$_2$. This prediction has been
experimentally verified for
monolayer~\cite{Nicholas:Energy,Nicholas:EnergyLoss,Pallecchi:Hot} and also
recently for bilayer graphene~\cite{Nicholas:Bilayer}. We note that in monolayer
and bilayer graphene, the acoustic phonon-limited resistivity due to unscreened
(screened) deformation potential coupling shows a $\rho\sim T^4$ ($\rho\sim
T^6$) dependence~\cite{Sarma:Acoustic,Sarma:Chirality}. Again, this is the same
as the situation in monolayer MoS$_2$ due to unscreened (screened) deformation
potential and piezoelectric interaction~\cite{Kaasbjerg:MoS2Acoustic}.

The $P\sim n^{-1/2}$ ($P_e\sim n^{-3/2}$) density dependence is same as in
conventional 2DEGs~\cite{Butcher:Thermoelectric} and in bilayer
graphene~\cite{Kubakaddi:Electronic} whereas it is different from monolayer
graphene where $P \sim n^{1/2}$ ($P_e\sim
n^{-1/2}$)~\cite{Kubakaddi:Interaction,Heikkila:ElectronPhonon}. In conventional
2DEGs, phonons are 3D and 2D electron dispersion is parabolic; in bilayer
graphene phonons are 2D and 2D electrons are with parabolic dispersion and in
monolayer graphene phonons are 2D and 2D electrons are with linear
dispersion. In view of this, we conclude that the difference in the dependence
on $n$ is due to the difference in the electronic density of states and
independent of phonon dimensionality.

From Eqs.~\eqref{eq:Sigma_TADP}--\eqref{eq:Sigma_PZ} valid in the BG regime, one
may compare the magnitude of the effective coupling constant $\Sigma$ comprising
all the acoustic el-ph couplings in MoS$_2$ to the one in monolayer and bilayer
graphene. For monolayer MoS$_2$ at $T, T_e \ll T_\text{BG}$ we find 
\begin{equation}
  \label{eq:SigmaMoS2}
\Sigma_\text{MoS$_2$}   \sim  4.7   \,\tilde{n}^{-1/2}~\mathrm{W/K}^4\,\mathrm{m}^2 ,
\end{equation}
which, because of strong carrier screening~\cite{Kaasbjerg:MoS2Acoustic}, is
entirely due to the unscreened TA deformation potential coupling. Using the
theory from Refs.~\onlinecite{Kubakaddi:Interaction,Kubakaddi:Electronic} for
the two graphene variants (with a deformation potential of $D=20$~eV and sound
velocity $c_s=2\times10^4$~m/s), we find 
\begin{align}
  \label{eq:monolayer}
  \Sigma_\text{monolayer} & \sim  0.058 \,\tilde{n}^{1/2}~\mathrm{W/K}^4\,\mathrm{m}^2 ,\\
  \label{eq:bilayer}                    
  \Sigma_\text{bilayer}   & \sim  0.075
  \,\tilde{n}^{-1/2}~\mathrm{W/K}^4\,\mathrm{m}^2 .
\end{align}
Thus, at low temperatures and $n=10^{12}$~cm$^{-2}$, the cooling power in
MoS$_2$ is almost two orders of magnitude larger than in mono- and bilayer
graphene. However, due to the density scaling of $\Sigma_\text{monolayer}$, the
difference in $P$ between monolayer MoS$_2$ and graphene decreases (increases)
at higher (lower) carrier densities, while it is independent on the density for
bilayer graphene. In addition, the differences in $P$ must be expected to
decrease at higher temperatures. This is due to the fact that the BG
temperatures in monolayer MoS$_2$ are lower, and hence, the transition to the EP
regime where the cooling power has a weaker temperature dependence $P\sim T$
(see below) takes place at lower temperatures in monolayer MoS$_2$ as compared
to mono- and bilayer graphene.

Unlike mobility, experimental measurements of $P$ have been useful to determine
the el-ph couplings, i.e. the deformation potentials, in conventional
2DEGs~\cite{Harris:EnergyLoss} and mono- and bilayer
graphene~\cite{Nicholas:Energy,Nicholas:EnergyLoss,Pallecchi:Hot,Schwab:Measurement,Nicholas:Bilayer}. The
results presented here apply to monolayers of MoS$_2$ and other transition metal
dichalcogenides where $\Sigma \sim \Xi_\lambda^2, e_{11}^2$ and may be helpful
to verify theoretically predicted coupling constants as in, e.g., monolayer
MoS$_2$~\cite{Kaasbjerg:MoS2,Kaasbjerg:MoS2Acoustic}.

\subsubsection{Analytic high-temperature limits}

In the high-temperature EP regime where $N_B(T_e) \sim k_\text{B} T_e /
\hbar\omega_{\lambda\bq}$, there are different relevant situations depending on
the degeneracy regime of the 2DEG and the screening of the el-ph interaction.

Starting with the deformation potential interaction, we find in the case of a
degenerate 2DEG ($T_e \ll T_F$) for the \emph{unscreened} coupling
\begin{equation}
  \label{eq:Sigma_TADP_high}
  \Sigma_\lambda^\text{DP} = \frac{2^2g_sg_v m^3 \Xi_\lambda^2 k_\text{B} E_F}{3\pi^2 \hbar^5\rho}
  \sim n \quad \text{and} \quad \delta = 1  ,
\end{equation}
and for the \emph{screened} coupling
\begin{equation}
  \Sigma_\lambda^\text{DP} = \frac{2^5g_sg_v m^4 \Xi_\lambda^2 k_\text{B} E_F^2}
                        {5\pi^2 \hbar^7\rho q_\text{TF}^2}
  \sim n^2 \quad \text{and} \quad \delta = 1  ,
\end{equation}
respectively. For a nondegenerate carrier distribution and neglecting the weak
Debye-H{\"u}ckel screening, we find
\begin{equation}
  \label{eq:Sigma_DP_highnon}
  \Sigma_\lambda^\text{DP} = \frac{2 m^2 \Xi_\lambda^2 k_\text{B}}{\hbar^3\rho} n
  \quad \text{and} \quad \delta = 1  .
\end{equation}
The $\delta = 1$ temperature dependence predicted in
Eqs.~\eqref{eq:Sigma_TADP_high}--\eqref{eq:Sigma_DP_highnon} is similar to 3D
bulk systems~\cite{SmithJensen}, conventional 2DEGs~\cite{Ridley:Hot} and
graphene~\cite{Sarma:EnergyRelaxation,Heikkila:ElectronPhonon} and originates
from phonon equipartition.

Due to the presence of the erfc in the matrix element for the piezoelectric
interaction~\eqref{eq:M_PE}, a simple analytic result cannot be obtained. For
unscreened piezoelectric scattering in the equipartition regime, we find
numerically that $\delta \lesssim 1$. The difference between the $\delta$ values
for deformation potential and piezoelectric coupling can be attributed to the
erfc in the matrix element for piezoelectric coupling.
\begin{figure}[!t]
  \includegraphics[width=0.49\linewidth]{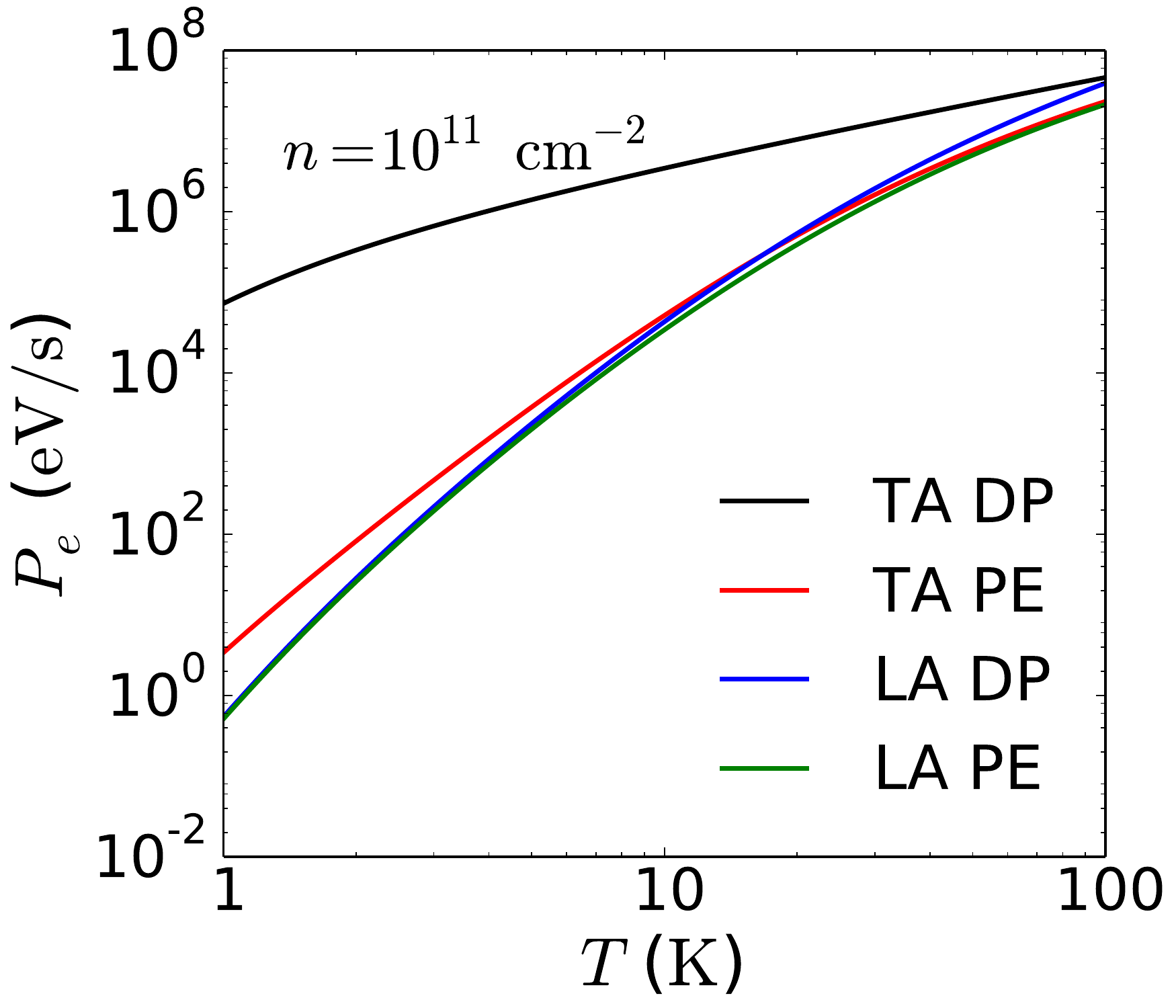}
  \includegraphics[width=0.49\linewidth]{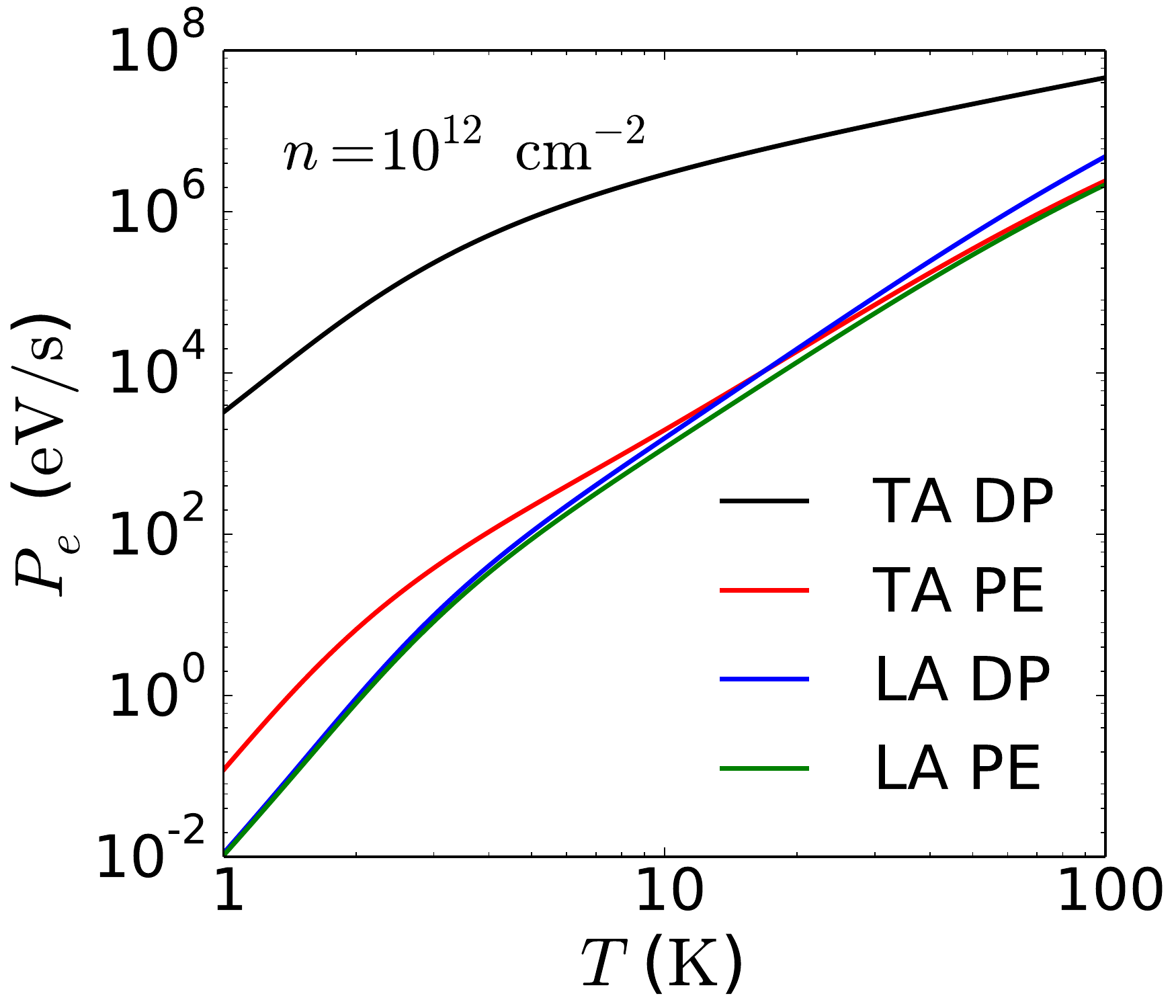}
  \caption{(Color online) Cooling power per electron due to the different
    coupling mechanisms to the acoustic phonons for $n=10^{11},
    10^{12}$~cm$^{-2}$ and lattice temperature $T=0$~K.}
\label{fig:P_vs_T_acoustic_individual}
\end{figure}

\subsubsection{Numerical results}

In the following, we present our numerical results for the temperature and
carrier density dependence of the cooling power due to acoustic phonon
scattering.

\begin{figure}[!t]
  \includegraphics[width=0.65\linewidth]{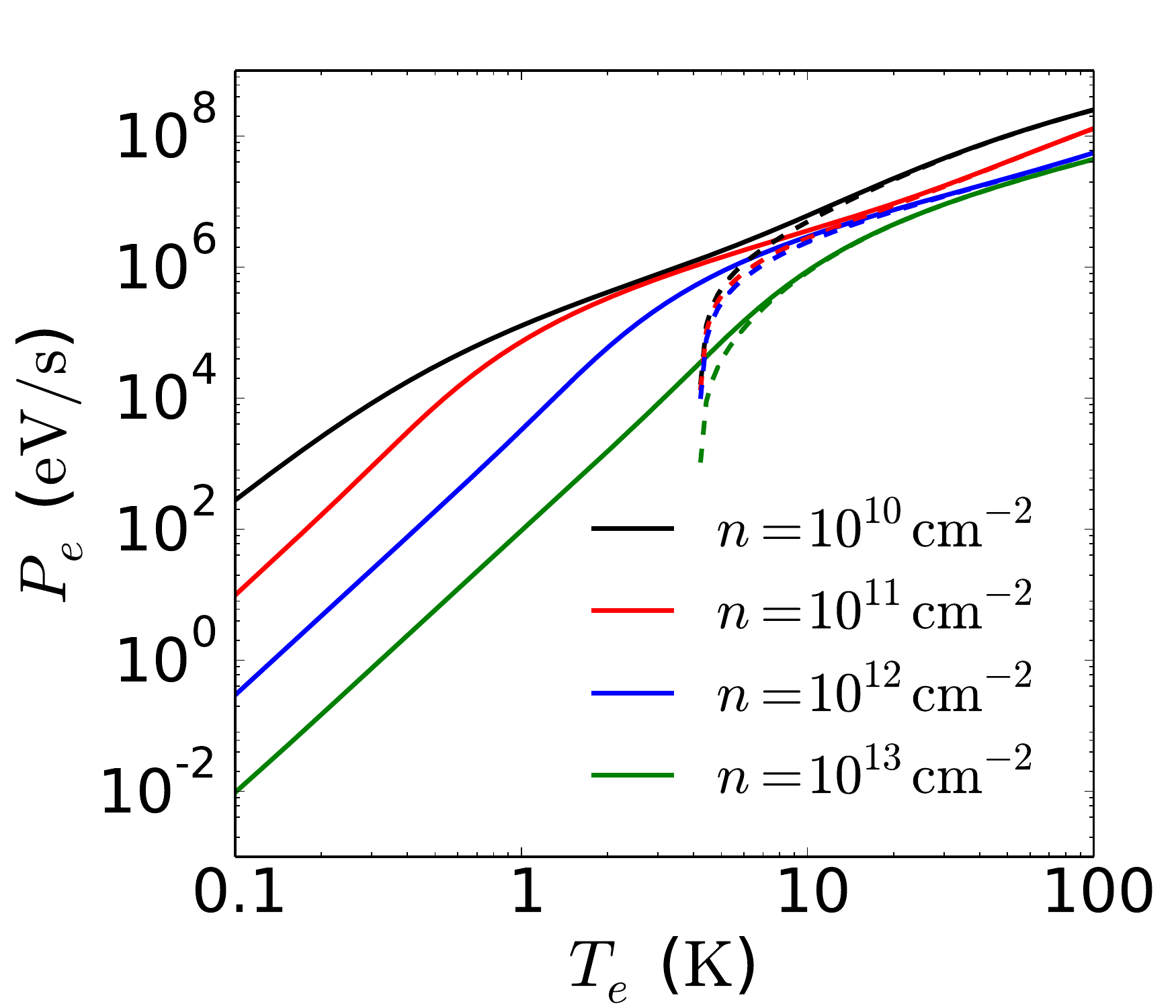}
  \caption{(Color online) Total cooling power per electron vs temperature due to
    acoustic phonon scattering at different carrier densities and lattice
    temperatures $T=0$~K (solid) and $T=4.2$~K (dashed).}
\label{fig:P_vs_T_acoustic}
\end{figure}

\begin{figure}[!t]
  \includegraphics[width=0.49\linewidth]{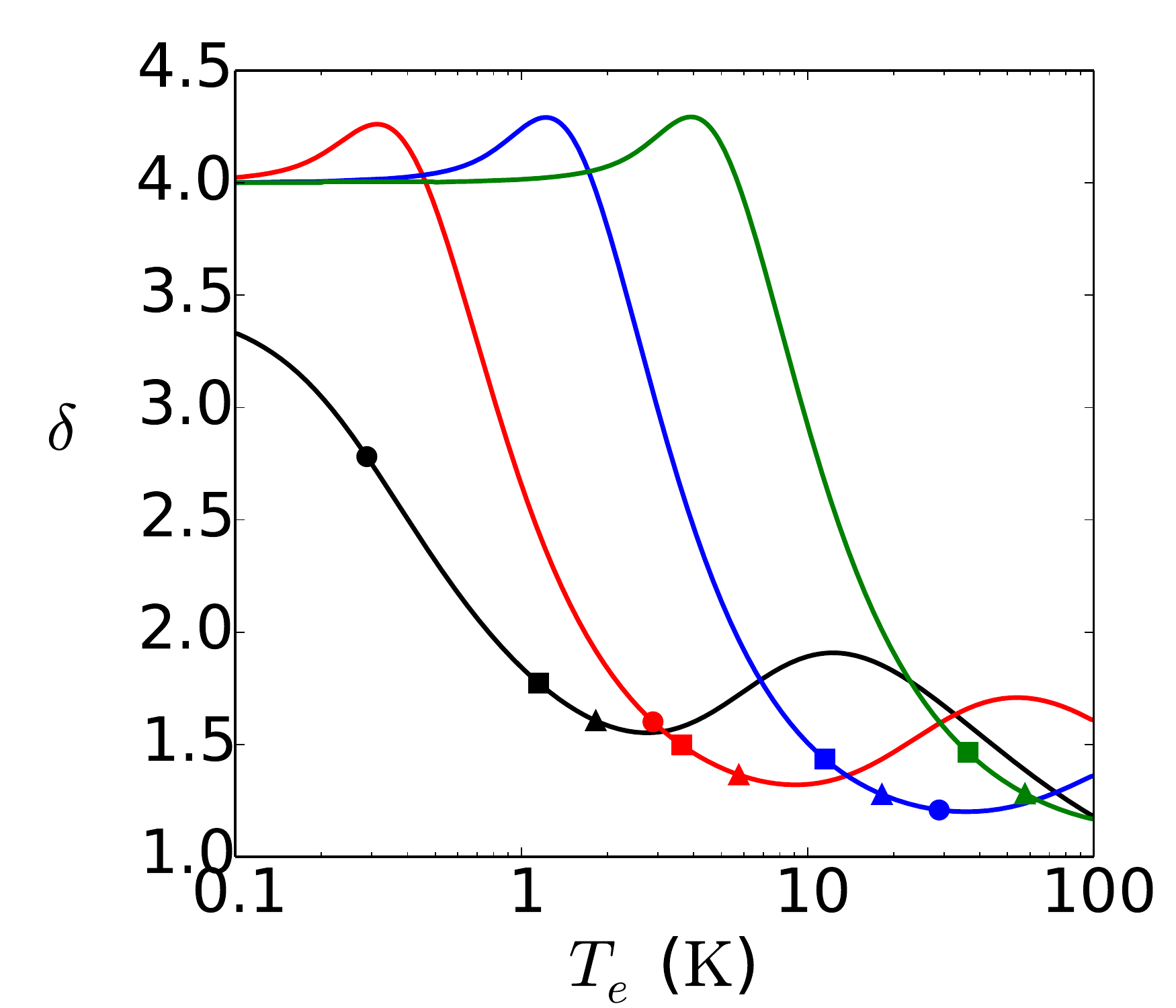}
  \includegraphics[width=0.49\linewidth]{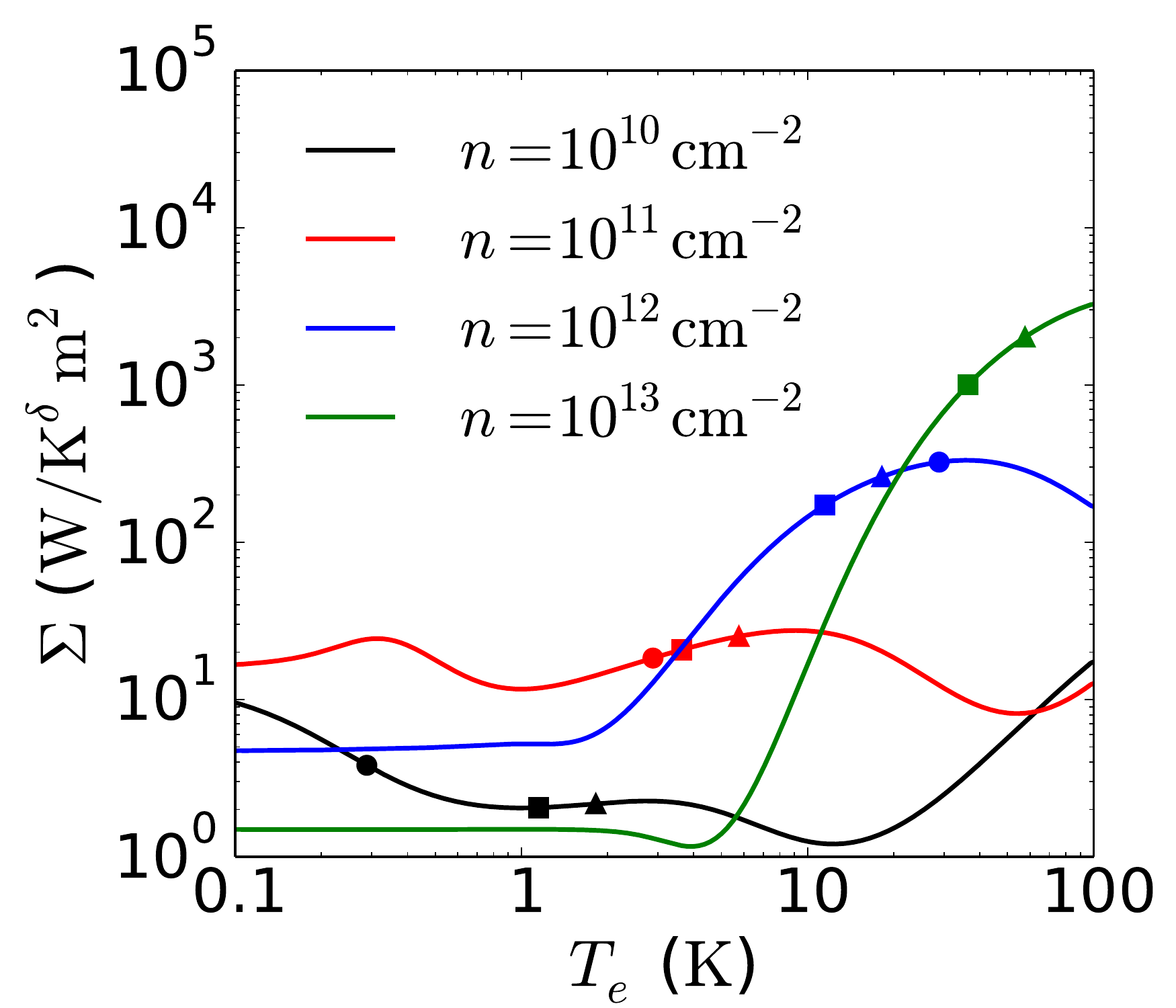}
  \caption{(Color online) Temperature dependence of the exponent $\delta$ and
    the effective coupling constant $\Sigma$ in the power-law expression for the
    cooling power in Eq.~\eqref{eq:P_acoustic} at different densities and
    lattice temperature $T=0$~K. The symbols mark the Fermi temperature
    $T_F=E_F/k_\text{B}$ ($\bullet$) and the BG temperatures for the TA
    ($\blacksquare$) and LA ($\blacktriangle$) phonons.}
\label{fig:effective}
\end{figure}

We start by discussing the dependence on the hot-electron temperature $T_e$ at
different carrier densities. The cooling power per electron for the different
coupling mechanisms is shown in Fig.~\ref{fig:P_vs_T_acoustic_individual}, while
Fig.~\ref{fig:P_vs_T_acoustic} shows the total cooling power due to all the
acoustic phonon coupling mechanisms for lattice temperatures $T=0,4.2$~K and
carrier densities $n =10^{10}-10^{13}$~cm$^{-2}$. The extracted values for the
exponent $\delta$ and the \emph{effective} coupling constant $\Sigma$ in
Eq.~\eqref{eq:P_acoustic} are shown in Fig.~\ref{fig:effective} for $T=0$.

From the individual contributions in Fig.~\ref{fig:P_vs_T_acoustic_individual},
the cooling power due to the unscreened deformation potential coupling to the TA
phonon is seen to dominate the other over the entire temperature range
considered. The same holds for the acoustic-phonon limited mobility and is due
to strong screening of the other acoustic el-ph
couplings~\cite{Kaasbjerg:MoS2Acoustic}. With increasing temperature, the
transition from Thomas-Fermi to Debye-H{\"u}ckel screening at $T_e \gg T_F$,
where $T_F$ is the Fermi temperature, results in a reduction of the screening
efficiency. At $n=10^{11}$~cm$^{-2}$ and $T_e\sim 100$~K screening is negligible
and the cooling power due to the different coupling mechanisms become
comparable.

The total cooling power due to acoustic phonon scattering is shown in
Fig.~\ref{fig:P_vs_T_acoustic} for lattice temperatures $T=0$ (full lines) and
$T=4.2$~K (dashed lines). At $T=0$~K, the cooling power approaches a $P \sim
T_e^4$ behavior in the BG regime due to the dominating unscreened deformation
potential coupling to the TA phonon. For finite lattice temperatures, the
cooling power vanishes at $T_e=T$. This gives rise to a significant drop in the
cooling power when $T_e$ approaches $T$ (see dashed lines). At $T_e < T$ the
electron distribution is heated by the lattice instead of cooled. At $T_e \gg
T$, the cooling power is dominated by the first term in
Eq.~\eqref{eq:P_acoustic} and the dashed lines merge with the full lines

The temperature dependence of the exponent $\delta$ and the \emph{effective}
coupling constant $\Sigma$ extracted from the calculated cooling power in
Fig.~\ref{fig:P_vs_T_acoustic} are shown in Fig.~\ref{fig:effective}. In the BG
regime, $T_e \ll T_\text{BG}$, all the curves, except the one for
$n=10^{10}$~cm$^{-2}$, saturate according to our analytic prediction for
unscreened deformation potential interaction in Eq.~\eqref{eq:Sigma_TADP},
i.e. $\delta=4$ and $\Sigma \sim n^{-1/2}$. For the lowest carrier density the
assumption $T_\text{BG} < T_F$ is not fulfilled, implying that the analytic
limit is not observed. For the largest carrier densities where $T_F \gtrsim
T_\text{BG}$, the strong temperature dependence of $\delta$ and $\Sigma$ at $T_e
\lesssim T_\text{BG}$ stems from the transition to the degenerate EP regime with
the limiting behavior for unscreened deformation potential interaction in
Eq.~\eqref{eq:Sigma_TADP_high}. At $T_e \gtrsim T_\text{BG}$, this gives rise to
a peak in $\Sigma$ with the maximum value given roughly by the limit in
Eq.~\eqref{eq:Sigma_TADP_high}, thus indicating that the degenerate EP limit is
a good approximation even at $T_e \sim T_\text{BG}$ and with $T_e > T_F$ for the
smallest densities. The nonmonotonic behavior of $\delta$ and $\Sigma$ at $T_e >
T_\text{BG}$ can be attributed to the temperature dependence of the screening
function. In the nondegenerate EP regime, $T_e \gg T_\text{BG}, T_F$, the
exponent approaches the analytic high-temperature limiting value $\delta\sim 1$.
\begin{figure}[!b]
  \includegraphics[width=0.65\linewidth]{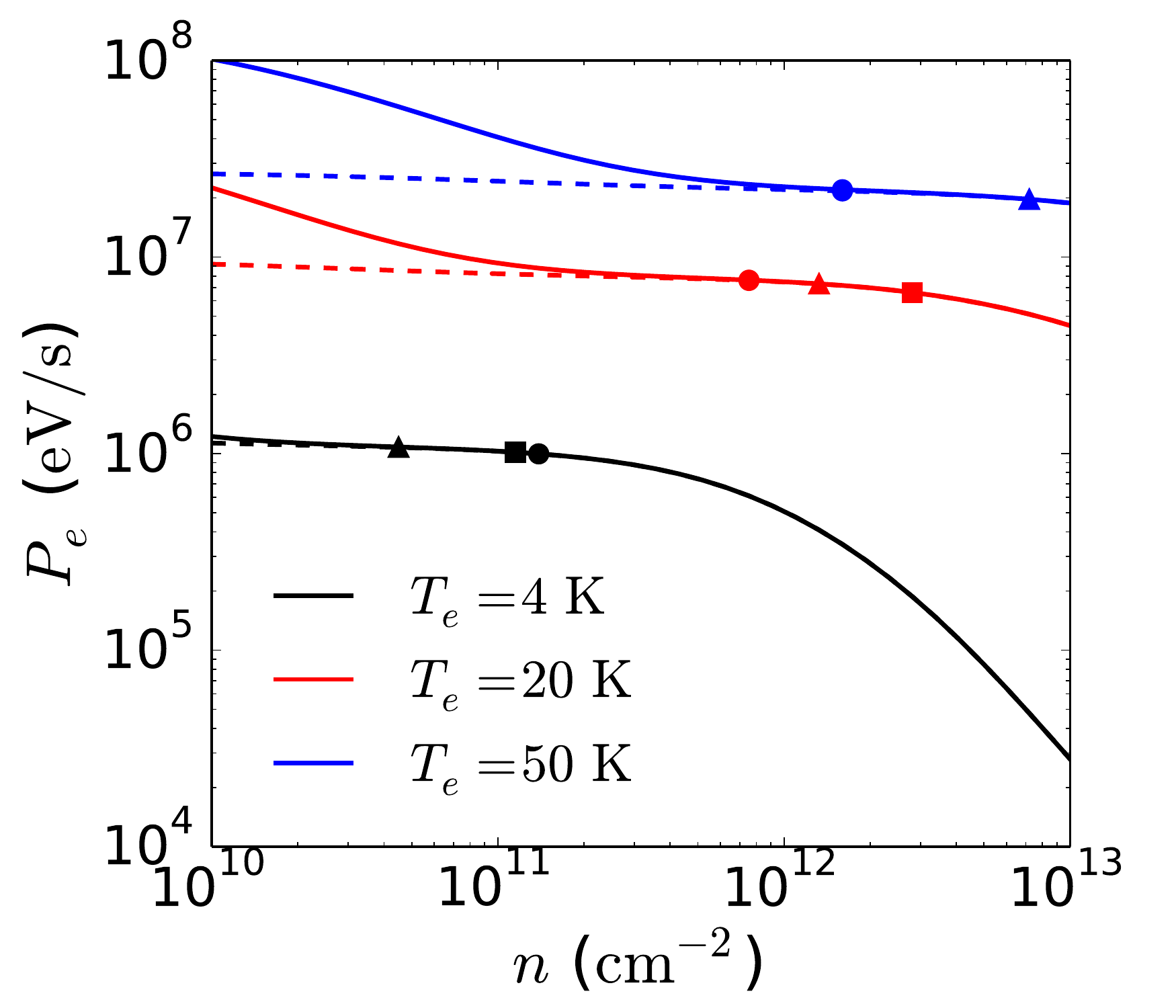}
  \caption{(Color online) Cooling power per electron vs carrier density for
    acoustic-phonon scattering (DP plus PE coupling to the TA and LA modes) at
    different electron temperatures (lattice temperature $T=0$~K). The dashed
    lines show the results with a background dielectric constant of $\kappa=5$
    corresponding to a dielectric with an intermediate $\kappa$ value. The
    symbols mark the densities where the Fermi temperature $T_F=E_F/k_\text{B}$
    ($\bullet$) and the BG temperatures for the TA ($\blacksquare$) and LA
    ($\blacktriangle$) phonons are equal the electron temperature $T_e$.}
\label{fig:P_vs_n_acoustic}
\end{figure}

The density dependence of the cooling power per electron is shown in
Fig.~\ref{fig:P_vs_n_acoustic} for $T_e = 4, 20, 50$~K at $T=0$~K. At the lowest
temperature $T_e = 4$~K, $P_e$ decreases with increasing $n$ and behaves as $P_e
\sim n^{-3/2}$ ($P\sim n^{-1/2}$) at high carrier densities in agreement with
the analytic limits in Eqs.~\eqref{eq:Sigma_TADP}--\eqref{eq:Sigma_PZ}. For the
two higher temperatures $T_e = 20$~K and 50~K, the density dependence of $P_e$
is weaker (stronger) at high (low) densities. At high densities, this is
due to a transition to the degenerate EP regime where $P_e \sim n^0$ ($P \sim
n$) for unscreened el-ph interaction (see Eq.~\eqref{eq:Sigma_TADP_high}). At
low densities, the observed $P_e \sim n^{\alpha}$ ($P \sim n^{\beta}$) behavior
with $\alpha \sim -0.5$--0 ($\beta \sim 0.5$--1) is only in partial agreement
with the analytic unscreened, high-temperature limit for nondegenerate carriers
in Eq.~\eqref{eq:Sigma_DP_highnon} and must hence be attributed to screening
effects. This interpretation is supported by the dashed lines in
Fig.~\ref{fig:P_vs_n_acoustic} which show $P_e$ in the presence of dielectric
background screening with $\kappa = 5$. The inclusion of background screening
implies that 2DEG screening becomes irrelevant at low densities and the
unscreened limit in Eq.~\eqref{eq:Sigma_DP_highnon} is realized.

\begin{figure}[!t]
  \includegraphics[width=0.49\linewidth]{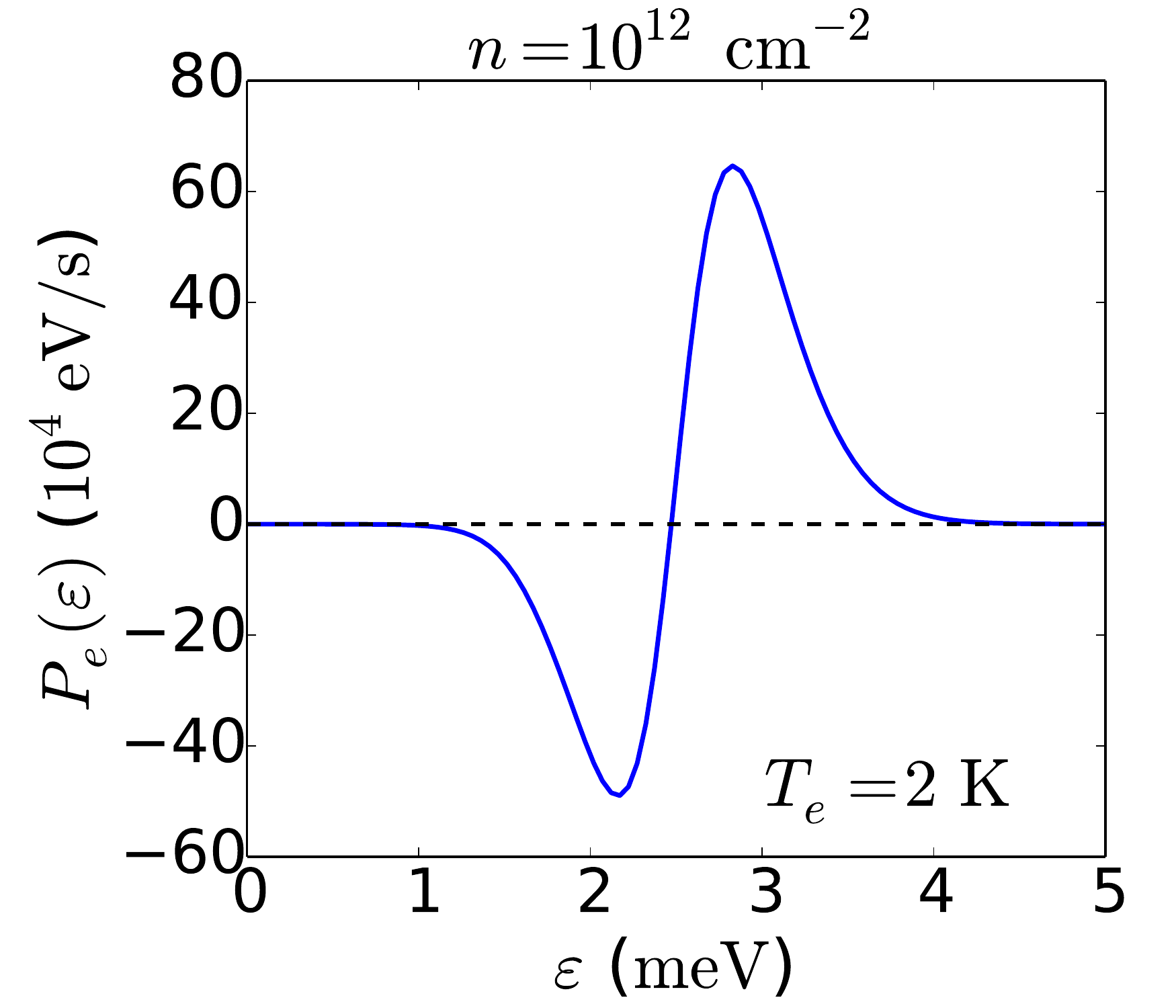}
  \includegraphics[width=0.49\linewidth]{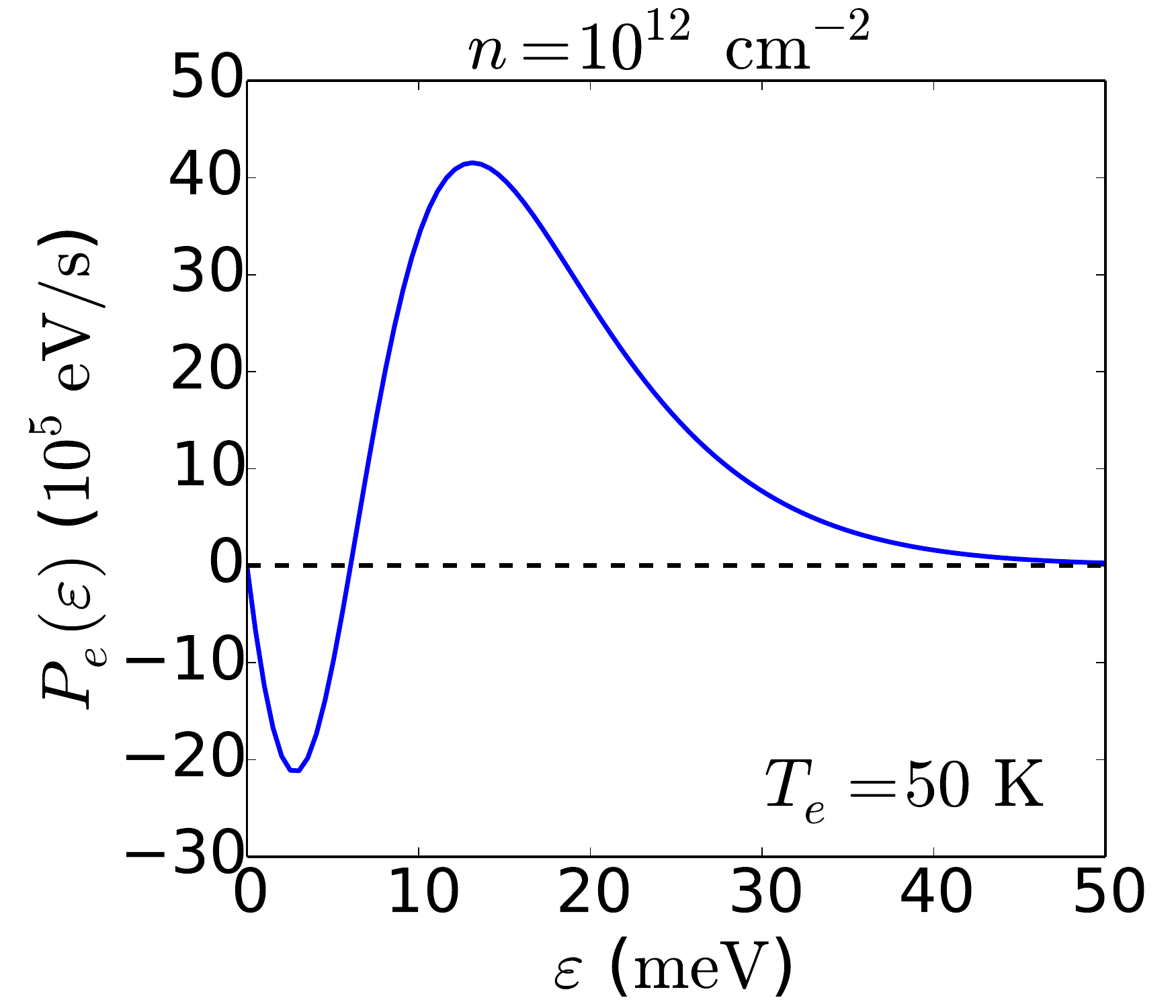}
  \caption{(Color online) Carrier-energy relaxation rate (Eq.~\eqref{eq:P_e}) vs
    carrier energy for acoustic-phonon scattering at a carrier density of
    $n=10^{12}$ cm$^{-2}$ and electron temperature $T_e=2$~K (left) and
    $T_e=50$~K (right). This corresponds to a degenerate gas in the BG regime
    and a non-degenerate gas in the high-temperature regime, respectively. The
    lattice temperature is $T=0$~K.}
\label{fig:P_vs_energy}
\end{figure}
We end this section by briefly discussing the carrier energy relaxation rate
which is shown in Fig.~\ref{fig:P_vs_energy} for a carrier density of
$n=10^{12}$ cm$^{-2}$ and temperatures $T=2$~K and $T=50$~K corresponding to a
degenerate and nondegenerate carrier distribution, respectively. For a
degenerate 2DEG, the transition from negative to positive energy relaxation rate
happens at $\varepsilon^*=E_F$. At carrier energies $\varepsilon_\bk < E_F$,
emission processes are Pauli blocked due to the filled Fermi sea and the states
have a net inflow of energy from higher energy states with
$\varepsilon_\bk>E_F$. In the nondegenerate regime, the Pauli blocking is
lifted. However, for low-energy states with band velocity $v_\bk < c_\lambda$,
simultaneous conservation of momentum and energy between initial and final state
is not possible for emission processes. Therefore, $P(\varepsilon)$ is
initially negative and decreasing with the carrier energy. The position of the
transition energy $\varepsilon^*$ is less obvious in the nondegenerate case and
will not be addressed in further detail here.
\begin{figure*}[!t]
  \centering
  \includegraphics[width=0.3\linewidth]{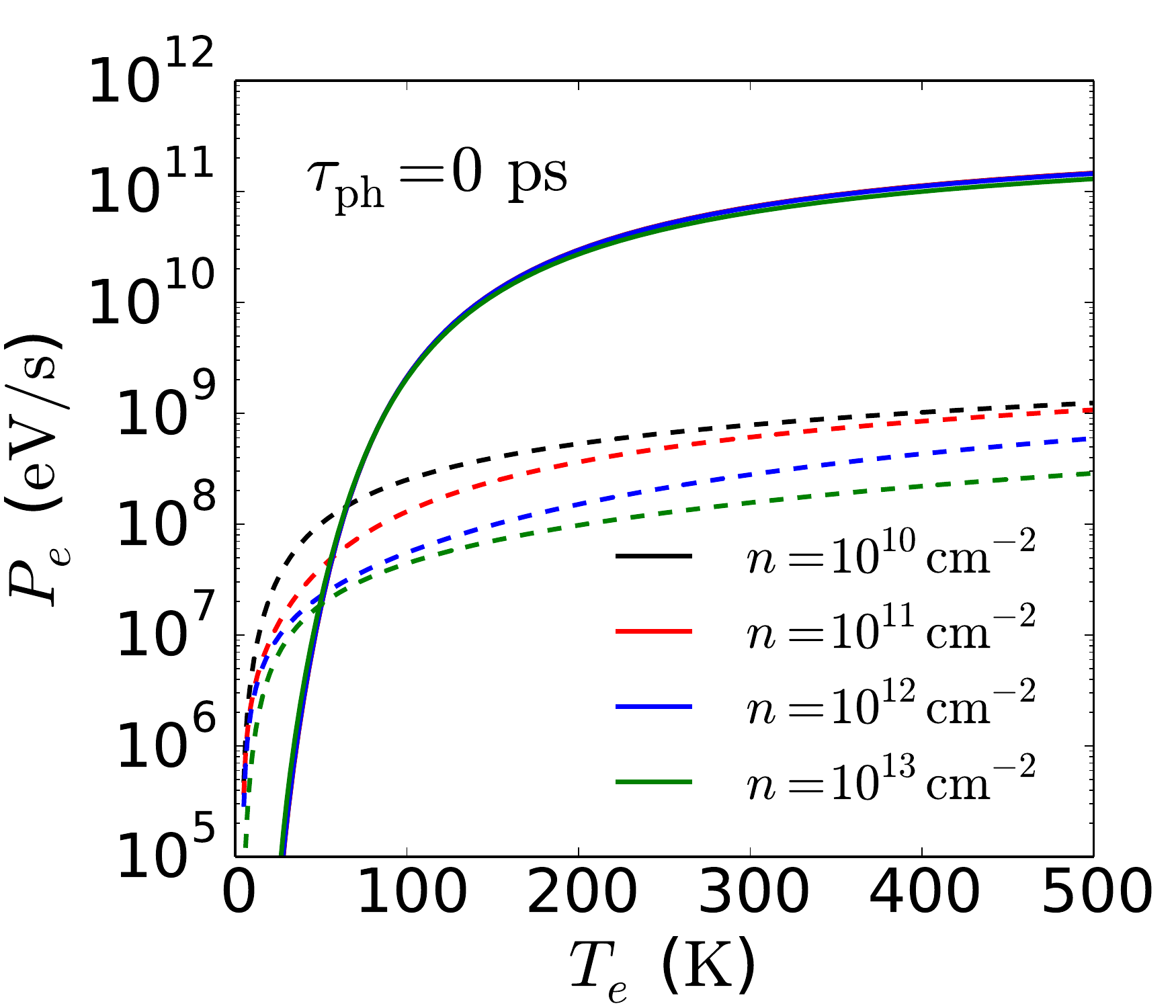}
  \includegraphics[width=0.3\linewidth]{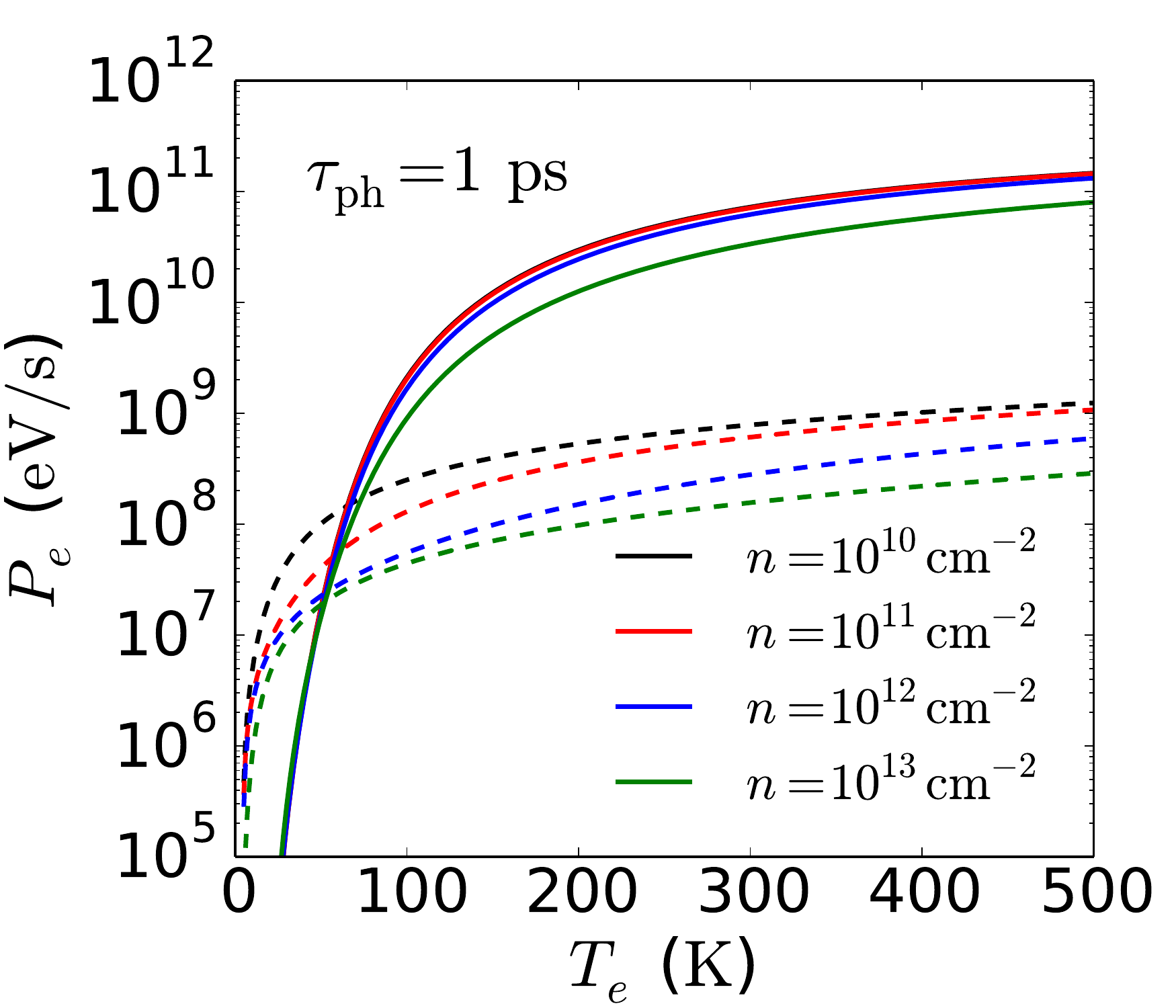}
  \includegraphics[width=0.3\linewidth]{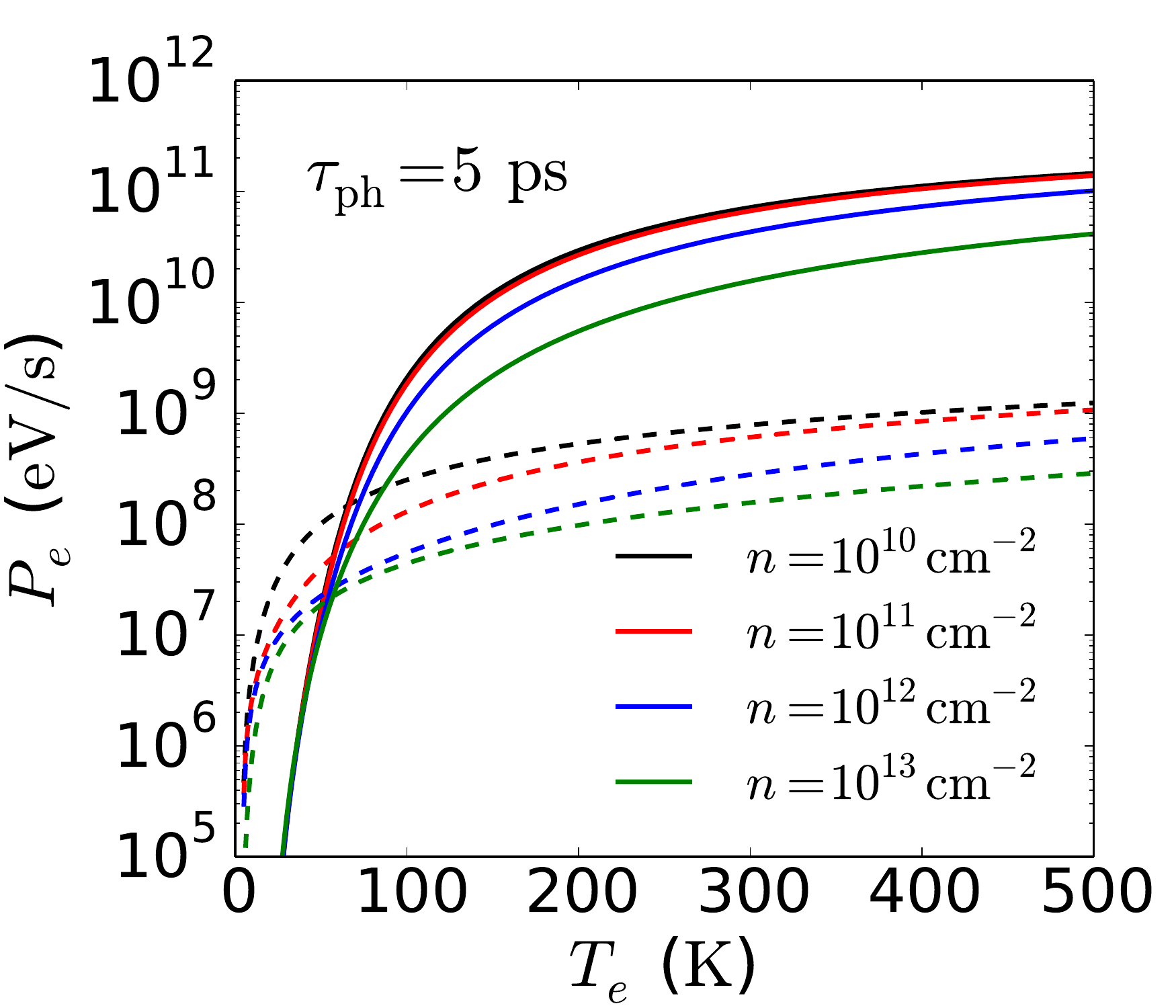}
  \caption{(Color online) Cooling power due to acoustic (dashed) and optical
    (full) phonons at different carrier densities and phonon lifetimes
    $\tau_\text{ph}=0,1, 5$~ps. The environmental temperature is $T=4.2$~K.}
  \label{fig:PvsT_acoustic_vs_optical}
\end{figure*}

\subsection{Cooling by acoustic and optical phonons at higher $T_e$}

In this final section, we consider the combined effect of acoustic and optical
phonon scattering on the cooling power. Estimates based on atomic
first-principles calculations of the optical phonon relaxation time due to
anharmonicity range from $\tau_\text{ph} \sim 1$--5~ps (corresponding to a
linewidth of $\gamma_\text{ph} \sim
1$--5~meV)~\cite{Zhang:Lattice,Mingo:Thermal}. Here, we present result for three
representative values $\tau_\text{ph}$ $= 0, 1, 5$~ps where
$\tau_\text{ph}=0$~ps corresponds to optical phonons in equilibrium with the
environment at temperature $T$.

In Fig.~\ref{fig:PvsT_acoustic_vs_optical} we show the cooling power due to both
acoustic and optical phonons as well as the individual contributions for
different carrier densities and phonon relaxation times. The crossover from
acoustic phonon to optical phonon dominated cooling power takes place in the
temperature interval $T_e \sim 50-75$~K depending on $\tau_\text{ph}$ and $n$.
The cooling power due to optical phonons is relatively independent of $n$ if the
hot-phonon effect is ignored, i.e. $\tau_\text{ph}=0$. With increasing
$\tau_\text{ph}$, the slower equilibration rate of the optical phonons gives
rise to phonon heating that leads to reabsorption processes and a decreasing
cooling power. Also it is observed that for a given $\tau_\text{ph}$, the
hot-phonon effect is larger for larger $n$, i.e. the reduction in cooling power
is larger. This behavior may be attributed to an increased scattering rate due
to the el-ph interaction for higher $n$ (see Fig.~\ref{fig:tauinv}). The
decrease in cooling power for higher $n$ is similar to the observations made in
bilayer graphene~\cite{Kubakaddi:Effect} and GaAs QWs~\cite{Huang:Comparison}
for surface-polar optical phonon scattering. For the largest values of
$\tau_\text{ph}$ ($=5$~ps) and $n$ ($=10^{13}$~cm$^{-2}$) chosen in the present
calculations, the hot-phonon effect reduces the cooling power to the optical
phonons by a factor of $\sim 3$. With decreasing temperature, the cooling power
due to optical phonons falls off as $\sim \exp(-\hbar\omega_\lambda / k_\text{B}
T_e)$ due to the exponential decaying occupation of electronic states with high
enough energy, $\varepsilon_\bk \gtrsim \hbar\omega_\lambda$, to emit an optical
phonon. Similar behavior for the cooling power due to optical phonons has been
demonstrated in graphene~\cite{Sarma:EnergyRelaxation}.

\begin{figure*}[!t]
  \centering
  \includegraphics[width=0.3\linewidth]{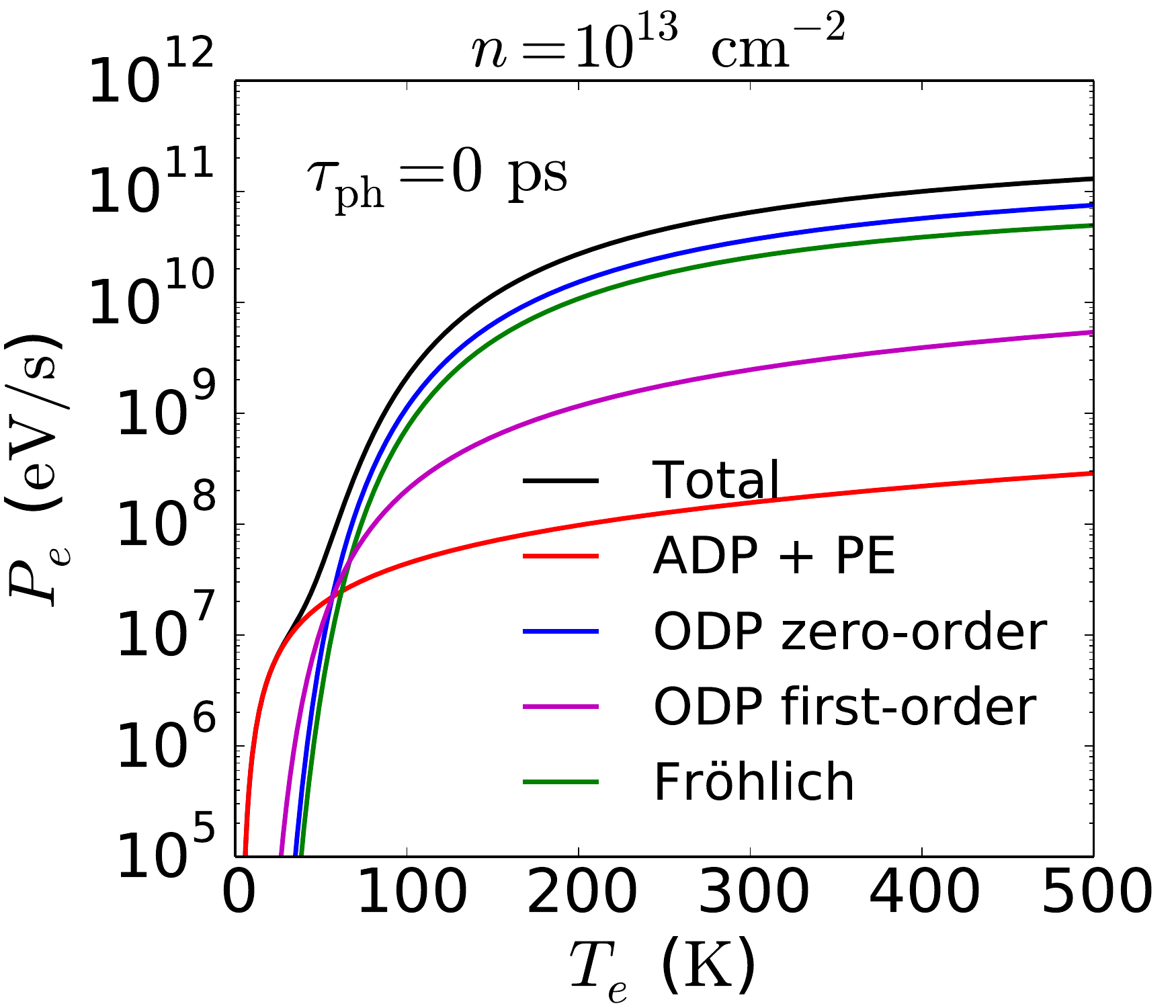}
  \includegraphics[width=0.3\linewidth]{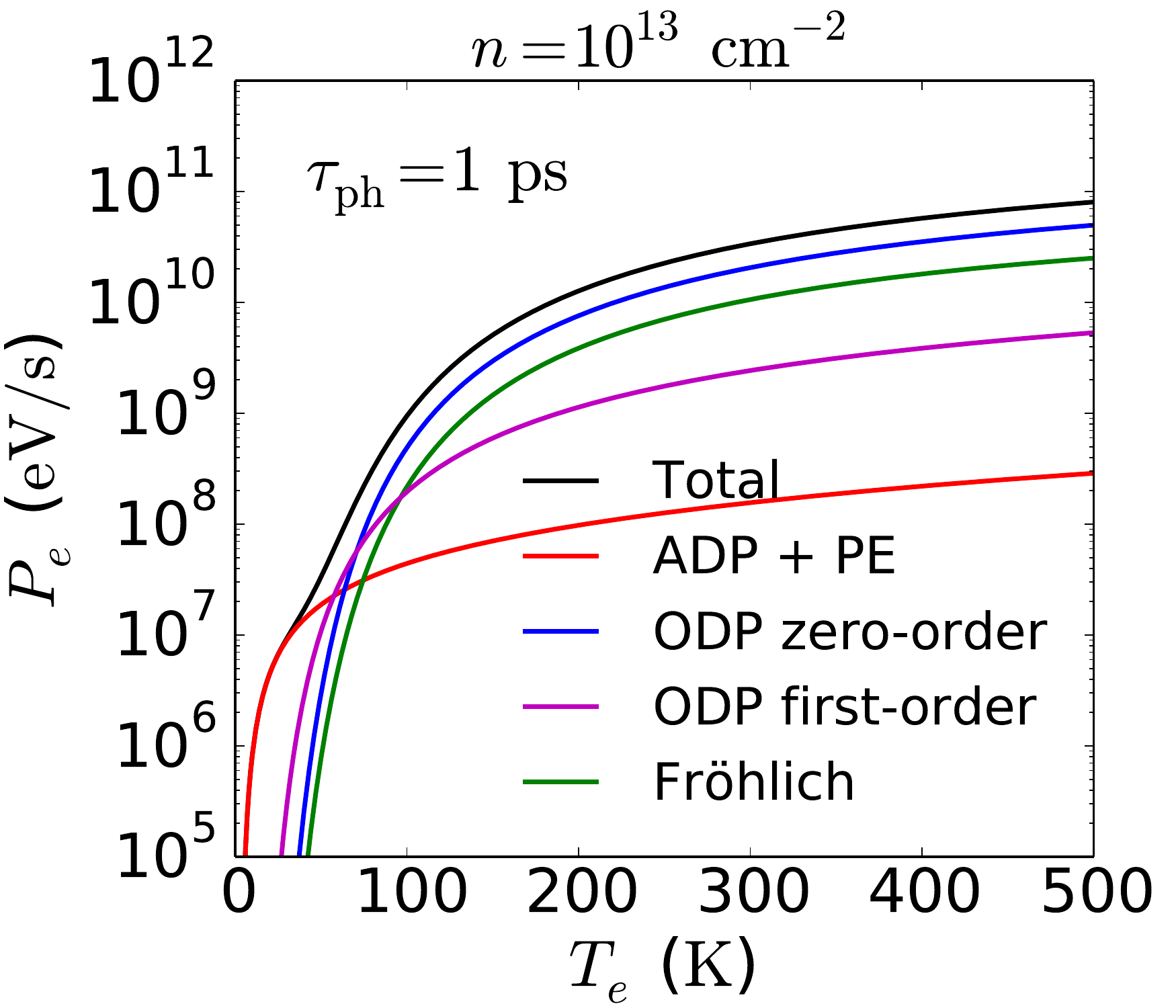}
  \includegraphics[width=0.3\linewidth]{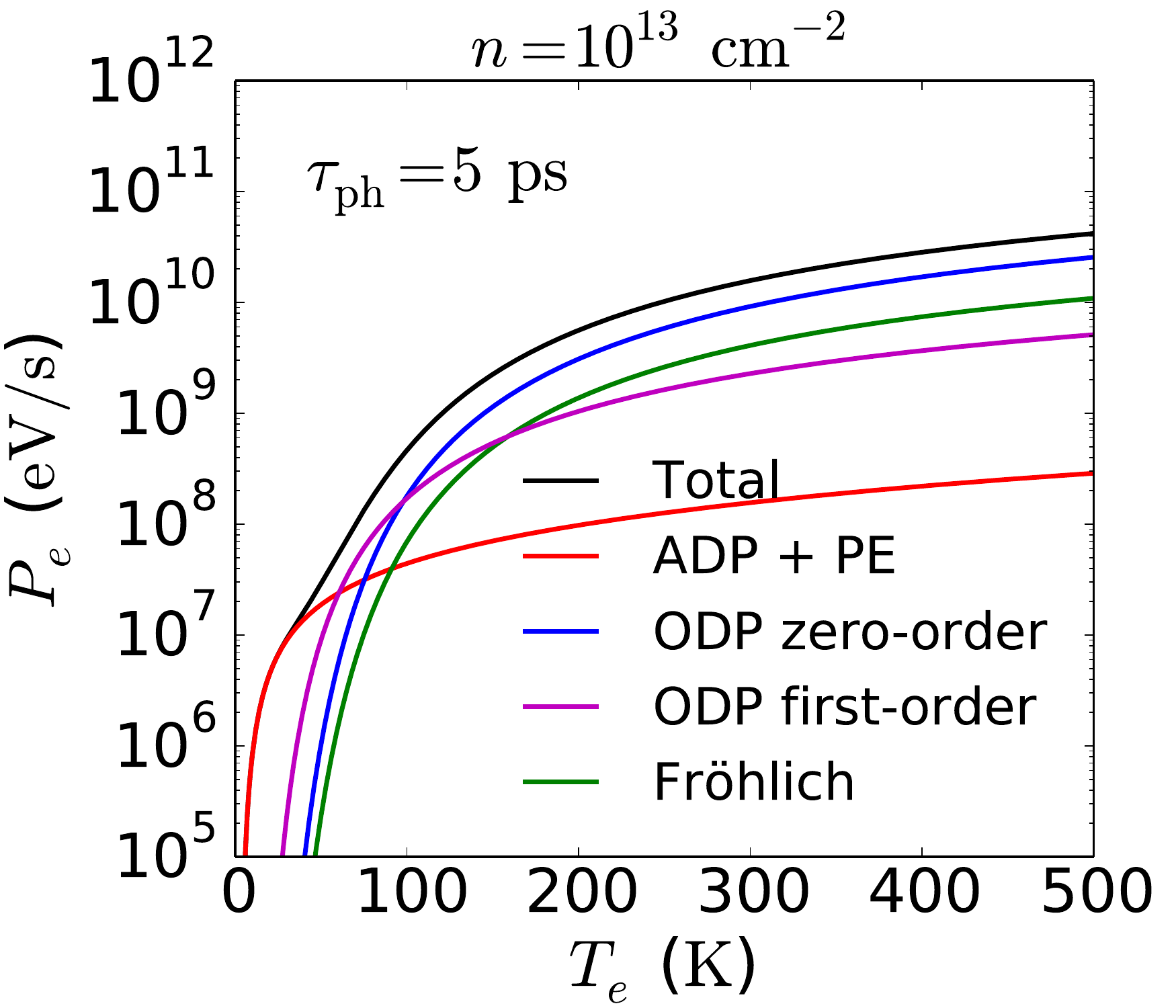}
  \caption{(Color online) Total cooling power per electron and the contributions
    from the individual coupling mechanisms to the acoustic and optical
    phonons. Results are shown for a carrier density of $n=10^{13}$~cm$^{-2}$
    and different phonon lifetimes $\tau_\text{ph}=0,1, 5$~ps. The environmental
    temperature is $T=4.2$~K.}
  \label{fig:PvsT_all}
\end{figure*}
In Fig.~\ref{fig:PvsT_all} the contributions to $P_e$ from the different
coupling mechanisms are shown. Overall, zero-order ODP and the Fr\"ohlich
interaction dominate the energy relaxation to the optical phonons. However, at
high carrier densities and large $\tau_\text{ph}$, the hot-phonon effect reduces
the cooling efficiency of the HP and LO phonons becoming comparable to that of
the phonons coupling via first-order ODP. The cooling power due to first-order
ODP does not change with increasing $\tau_\text{ph}$. This is due to the fact
that optical phonons coupling via first-order ODP do not heat up because of the
weak interaction and are therefore not subject to the hot-phonon effect.

\subsubsection{Heating of optical phonons}

\begin{figure}[!b]
  \centering
  \includegraphics[width=0.48\linewidth]{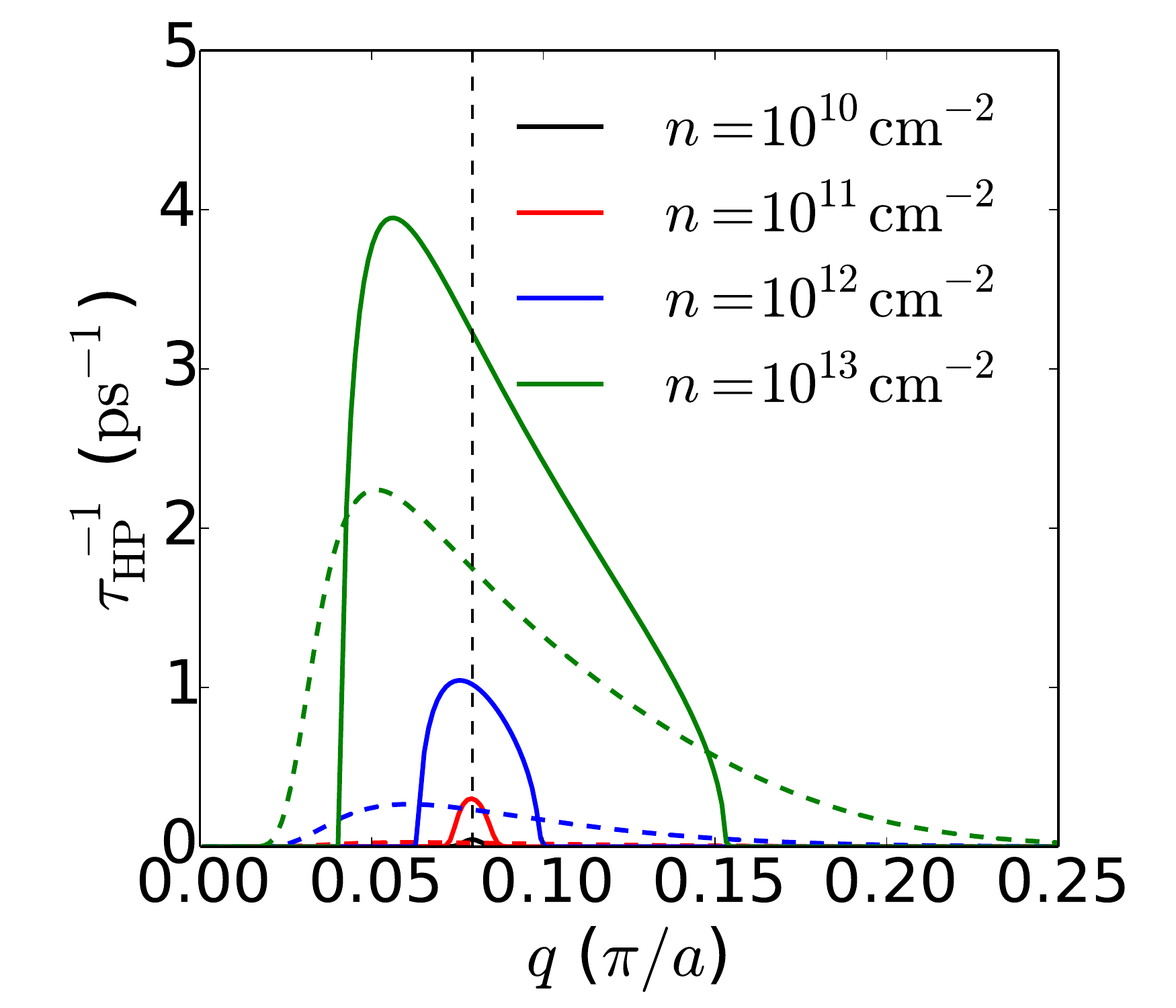}
  \includegraphics[width=0.48\linewidth]{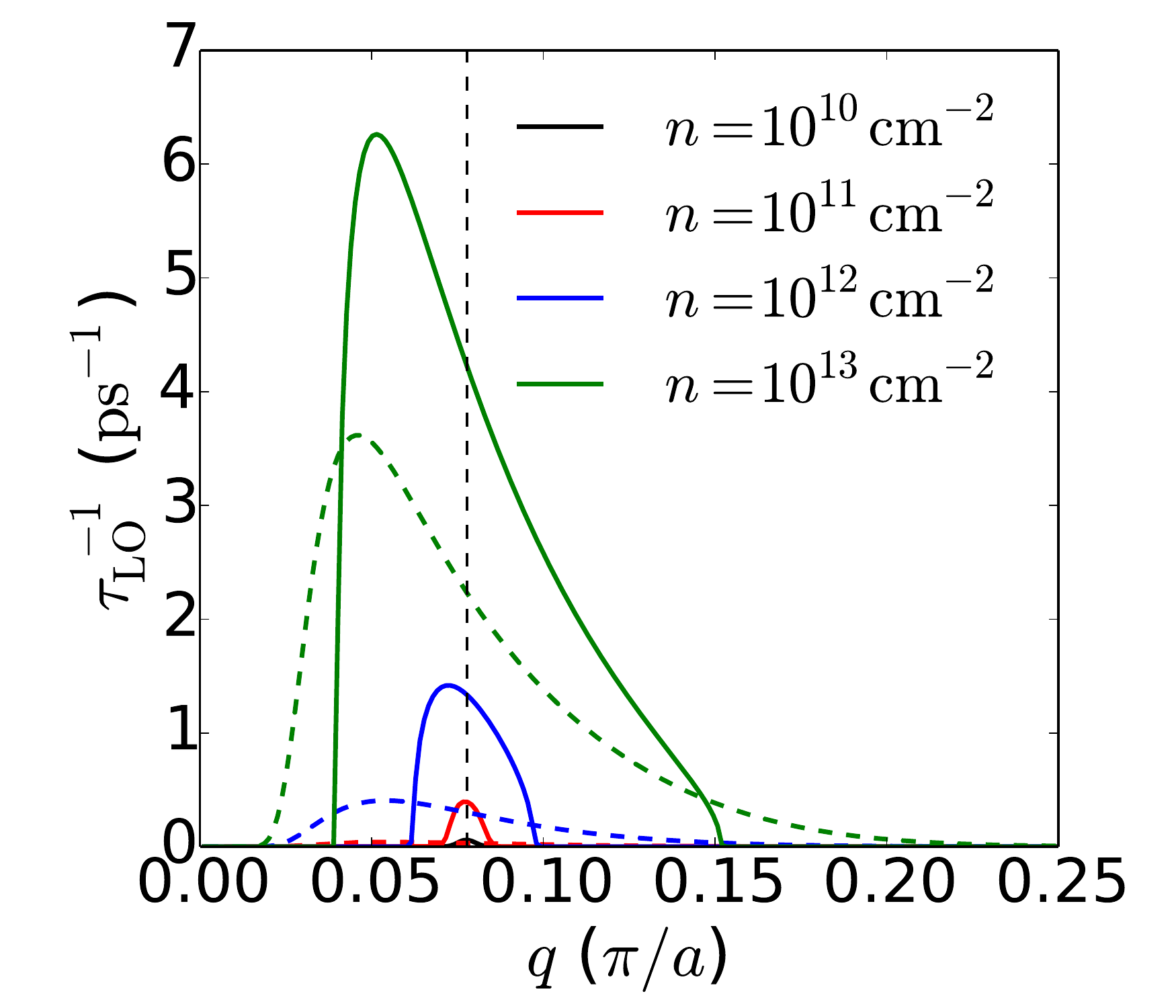}
  \caption{(Color online) Inverse phonon lifetime due to el-ph scattering for
    the HP (left) and LO (right) modes as a function of the phonon wave vector
    $q$ at $T_e=1$~K (full) and $T_e=300$~K (dashed). The vertical dashed lines
    mark the phonon wave vector $q=(2m\omega_\lambda / \hbar)^{1/2}$
    corresponding to electron-hole pair excitations between filled states at the
    bottom of the $K,K'$ valleys and empty states at $\varepsilon_\bk =
    \hbar\omega_\lambda$.}
  \label{fig:tauinv}
\end{figure}

In order to further analyze the heating of the optical phonons, we show in
Fig.~\ref{fig:tauinv} the inverse phonon lifetime $\tau_{\lambda\bq}^{-1}$ due
to el-ph scattering for the LO and HP phonons near the zone-center at different
carrier densities. It is important to note that only phonons with wave vectors
in a limited interval centered around $q = (2m\omega_\lambda / \hbar)^{1/2}$ are
subject to el-ph scattering. This value of $q$ (marked with the vertical dashed
lines in Fig.~\ref{fig:tauinv}) corresponds to intra-valley electron-hole pair
excitations between filled states at the bottom of the valley and empty states
at energy $\varepsilon_\bk = \hbar\omega_\lambda$. At low carrier densities and
low $T_e$ where only electronic states with energy $\varepsilon_\bk \ll
\hbar\omega_\lambda$ are occupied, electronic damping of the optical phonons is
only possible through these electron-hole pair excitations. The scattering rate
therefore becomes strongly peaked around this special $q$ value. At higher $n$
and $T_e$ where more phase space becomes available for el-ph scattering, the
peak broadens and the peak value is shifted to lower values of $q$.

The phonon linewidth due to el-ph scattering, $\gamma_{\lambda\bq} =
\hbar/\tau_{\lambda\bq}$, given by the inverse phonon lifetime in
Fig.~\ref{fig:tauinv} increases significantly with increasing carrier density
and becomes comparable to $\gamma_\text{ph}$ due to ph-ph scattering at the
highest carrier densities. Such a pronounced density dependence of the phonon
linewidth (and the accompanying frequency shift; see App.~\ref{sec:GF}) should
be observable in spectroscopy (Raman, x-ray or neutron) on gated samples where
the level of electron doping can be tuned. Indeed, we note that such an effect
has been observed experimentally in Raman spectroscopy on monolayer
MoS$_2$~\cite{Sood:Symmetry}, however, only the doping dependence of
$\Gamma$-point phonons where the effect is comparatively small was addressed.

\begin{figure}[!t]
  \centering
  \includegraphics[width=0.65\linewidth]{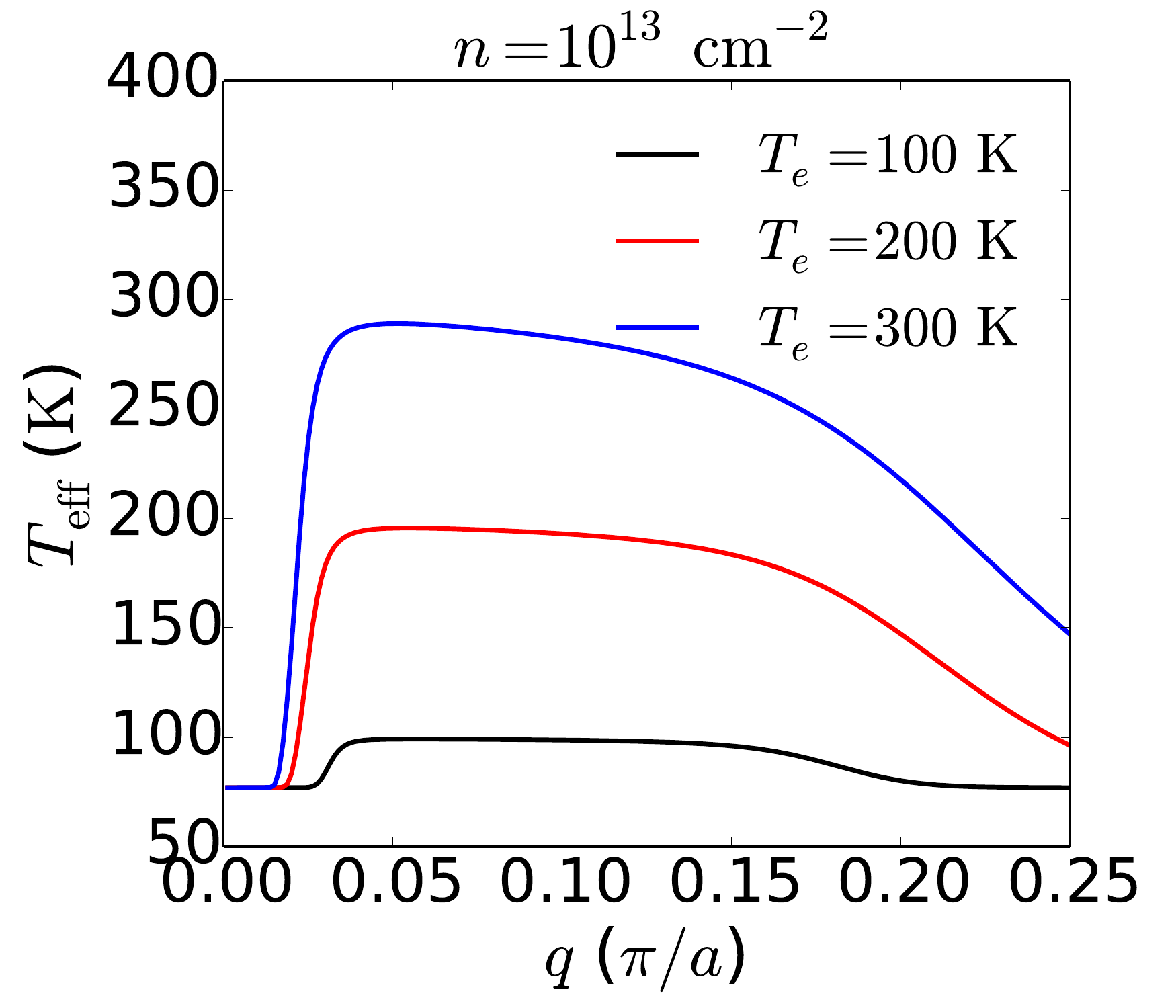}
  \caption{(Color online) Effective temperature for the optical HP phonon at
    carrier density $n=10^{13}$~cm$^{-2}$, environmental temperature $T=77$~K,
    ph-ph scattering lifetime $\tau_\text{ph}=5$~ps, and different values of the
    hot-electron temperature $T_e$.}
  \label{fig:Teff}
\end{figure}

Finally, in Fig.~\ref{fig:Teff} we show the effective hot-phonon temperature
$T_\text{eff}$ for the HP mode at high carrier density $n = 10^{13}$~cm$^{-2}$,
$\tau_\text{ph}=5$~ps, lattice temperature $T = 77$~K, and different electron
temperatures. $T_\text{eff}$, and hence the hot-phonon population
$N_{\lambda\bq}$, varies significantly with $q$. As anticipated, $T_\text{eff}$
is larger for larger $T_e$ and the $q$ dependence is determined by the inverse
phonon lifetime due to el-ph scattering in Fig.~\ref{fig:tauinv} through
Eq.~\eqref{eq:Nhot} for the hot-phonon distribution function. Each curve has
broad maximum in the neighborhood of the phonon wave vector $q =
(2m\omega_\lambda / \hbar)^{1/2}$ where phonon heating is most significant. This
is similar to findings in GaAs QWs~\cite{Huang:Comparison,Morkoc:QW} and in
bilayer graphene~\cite{Kubakaddi:Effect}. In the $q$ range where the el-ph
scattering rate in Fig.~\ref{fig:tauinv} exceeds $\tau_\text{ph}^{-1}$, the
effective temperature of the hot phonons approaches a value $T_\text{eff} \sim
T_e$ close to the hot-electron temperature which leads to the reduction of the
cooling power due to optical phonons shown in
Figs.~\ref{fig:PvsT_acoustic_vs_optical} and~\ref{fig:PvsT_all}.

\section{Conclusions}

Considering electron scattering from acoustic and optical phonons, we have
studied the electron-temperature $T_e$ and carrier density $n$ dependence of the
hot-electron cooling power $P$ in $n$-type monolayer MoS$_2$.  At low electron
temperatures $T_e < 50$--75~K, the cooling power is governed by scattering off
acoustic phonons with the unscreened DP coupling to the TA phonon dominating the
other contributions. In the Bloch-Gr{\"u}neisen regime, the unscreened DP
coupling shows a $P\sim T_e^4$ and $P\sim n^{-1/2}$ dependence. The cooling
power due to the screened DP coupling and screened PE coupling show $P\sim
T_e^6$ and $P\sim n^{-1/2}$ dependencies. These predicted $T_e$ dependencies are
characteristics of two-dimensional acoustic phonons. For higher temperatures
$T_e \gtrsim T_\text{BG}$, the exponent of $T_e$ gradually changes to lower
values and approaches $\delta \sim 1$ in the high-$T$ EP regime $T_e\gg
T_\text{BG},T_F$. In the extreme BG regime $T_e \ll T_\text{BG}$, the effective
coupling constant $\Sigma$ in Eq.~\eqref{eq:P_acoustic} saturates at a constant
density-dependent value which is almost two orders of magnitude larger than the
low-$T$ value of $\Sigma$ in mono- and bilayer graphene for
$n=10^{12}$~cm$^{-2}$. At higher temperatures, $\Sigma$ shows a nontrivial $T_e$
dependence.

The cooling power due to optical phonons (taking into account phonon heating)
dominates for $T_e \gtrsim 50$-75~K. The optical zero-order deformation
potential interactions and the Fr{\"o}hlich interaction to the LO phonon
dominate $P$ due to optical phonons. The hot-phonon effect is found to reduce
$P$ due to optical phonons by a factor $\sim 3$. The hot-phonon effect becomes
more significant at high values of $n$, $T_e$ and phonon relaxation time
$\tau_\text{ph}$, where the effective hot-phonon temperature reaches $T_e$ for
phonons with wave vectors in the neighborhood of $q = (2m\omega_\lambda /
\hbar)^{1/2}$. For low electron temperatures, $ k_\text{B} T_e \ll
\hbar\omega_{\lambda\bq}$, the cooling power due to optical phonons decreases
exponentially. Low-temperature experiments may validate the present predictions
for the temperature and carrier density dependence of the hot-electron cooling
power.

\begin{acknowledgments}
  K.K. acknowledges support from the Carlsberg Foundation.
\end{acknowledgments}

\appendix 

\section{Nonequilibrium Green function approach to phonon heating}
\label{sec:GF}

In this appendix we demonstrate the equivalence between the Boltzmann treatment
of phonon heating in Sec.~\ref{sec:hot} of the main part of the paper and a
quantum-kinetic description within the framework of the Keldysh nonequilibrium
Green function formalism~\cite{Jauho}.

\subsection{Phonon Green function}

In the presence of interactions, the retarded phonon Green function (GF) is
given by the Dyson equation $D_{\lambda\bq}^r(\omega)^{-1} =
D_{0,\lambda\bq}^r(\omega)^{-1} - \Pi_{\lambda\bq}^r(\omega)$ where
$D_{0,\lambda\bq}^r(\omega)=\tfrac{1}{\omega - \omega_{\lambda\bq} + i0^+} -
\tfrac{1}{\omega + \omega_{\lambda\bq} +i0^+}$ is the \emph{bare} phonon GF and
$\Pi_{\lambda\bq}^r(\omega)$ is the phonon self-energy~\cite{Mahan}. Neglecting
the small renormalization of the phonon frequencies due to the real part of
self-energy, one gets
\begin{align}
  \label{eq:GF}
  D_{\lambda\bq}^r(\omega) & = \frac{2\omega_{\lambda\bq}}
      {\omega^2 - \omega_{\lambda\bq}^2 - 2\omega_{\lambda\bq} \Pi_{\lambda\bq}^r(\omega)}
      \nonumber \\
      & = \frac{2\omega_{\lambda\bq}}
        {\omega^2 - \omega_{\lambda\bq}^2 + i\omega_{\lambda\bq}\gamma_{\lambda\bq}(\omega)}
\end{align}
where $\gamma_{\lambda\bq}(\omega) \equiv - 2\text{Im}\,
\Pi_{\lambda\bq}^r(\omega)$ is the damping function.

When the phonon linewidth is much smaller than the frequency
$\gamma_{\lambda\bq}(\omega_{\lambda\bq}) \ll \omega_{\lambda\bq}$, the phonon
GF can be approximated in the vicinity of the frequency $\pm
\omega_{\lambda\bq}$ by
\begin{equation}
  \label{eq:B}
  D_{\lambda\bq}^r(\omega) \approx 
     \frac{1}{\omega - \omega_{\lambda\bq} + i\gamma_{\lambda\bq}/2} - 
     \frac{1}{\omega + \omega_{\lambda\bq} +i\gamma_{\lambda\bq}/2} ,
\end{equation}
where $\gamma_{\lambda\bq} =
\gamma_{\lambda\bq}(\omega)\vert_{\omega=\omega_{\lambda\bq}}$. The
corresponding spectral function $B_{\lambda\bq}(\omega) = -2 \text{Im}\,
D_{\lambda\bq}(\omega)$ is given by two Lorentzians of width
$\gamma_{\lambda\bq}$ centered at the frequencies $\pm\omega_{\lambda\bq}$,
\begin{equation}
  B_{\lambda\bq}(\omega)
  = \frac{\gamma_{\lambda\bq}}
         {(\omega - \omega_{\lambda\bq})^2 + (\gamma_{\lambda\bq}/2)^2} - 
    \frac{\gamma_{\lambda\bq}}
         {(\omega + \omega_{\lambda\bq})^2 + (\gamma_{\lambda\bq}/2)^2} ,
\end{equation}
allowing us to identify the inverse phonon lifetime as
$\tau_{\lambda\bq}^{-1} = \gamma_{\lambda\bq}/\hbar$.

\subsubsection{Hot-phonon distribution function}

Due to the out-of-equilibrium situation, the lesser phonon GF must be obtained
from its Keldysh equation, $D_{\lambda\bq}^<(\omega) = D_{\lambda\bq}^r(\omega)
\Pi_{\lambda\bq}^<(\omega) D_{\lambda\bq}^a(\omega)$, which gives
\begin{equation}
  D_{\lambda\bq}^<(\omega) = - B_{\lambda\bq}(\omega) 
      \frac{\Pi_{\lambda\bq}^<(\omega)}{2\text{Im}\,\Pi_{\lambda\bq}^r(\omega)}. 
\end{equation}
In the limit where $\gamma_{\lambda\bq} \ll \omega_{\lambda\bq}$, the
Lorentzians in the spectral function~\eqref{eq:B} can be approximated by
$\delta$ functions implying that the lesser function can be written on the
quasi-equilibrium form
\begin{equation}
  D_{\lambda\bq}^<(\omega) \approx -i 
      \left[
        \delta(\omega + \omega_{\lambda\bq}) \left(1 + N_{\lambda\bq} \right)
        + \delta(\omega - \omega_{\lambda\bq}) N_{\lambda\bq}
      \right]
\end{equation}
where 
\begin{equation}
  \label{eq:Nnoneq}
  N_{\lambda\bq} = 
  \frac{i\Pi_{\lambda\bq}^<} {2\abs{\text{Im}\,\Pi_{\lambda\bq}^r}}
  \bigg\vert_{\omega=\omega_{\lambda\bq}}
\end{equation}
is the out-of-equilibrium phonon distribution function.

In the presence of coupling to an environmental phonon bath as well as the el-ph
interaction, the self-energy is given by the sum of the two contributions,
$\Pi_{\lambda\bq} = \Pi_\text{ph} + \Pi_{\lambda\bq}^\text{el-ph}$. The
imaginary parts of the two retarded self-energies are related to the respective
damping rates as $\gamma_\text{ph} =
-2\text{Im}\,\Pi_\text{ph}^r\vert_{\omega=\omega_{\lambda\bq}}$ and
$\gamma_{\lambda\bq}^\text{el-ph} = -2\text{Im} \,
\Pi_{\lambda\bq}^\text{el-ph,r}\vert_{\omega=\omega_{\lambda\bq}}$, where an
expression for the latter is given in Eq.~\eqref{eq:tau_phonon} below. The
lesser self-energy due to the coupling to environmental phonons at temperature
$T$ is given by $\Pi_\text{ph}^<\vert_{\omega=\omega_{\lambda\bq}} = -i N_B(T)
\gamma_\text{ph}$~\cite{Watanabe:Noneq}. The lesser self-energy due to the el-ph
interaction can be written on a similar form
$\Pi_{\lambda\bq}^{\text{el-ph},<}\vert_{\omega=\omega_{\lambda\bq}} = -i
N_B(T_e) \gamma_{\lambda\bq}^\text{el-ph}$, however, with the environmental
temperature replaced by the hot-electron temperature $T_e$ (this follows from
Eq.~\eqref{eq:tau_phonon} below). For the out-of-equilibrium distribution
function we thus obtain
\begin{equation}
  N_{\lambda\bq} = 
  \frac{\gamma_\text{ph} N_B(T) + \gamma_{\lambda\bq} N_B(T_e)}
       {\gamma_\text{ph} + \gamma_{\lambda\bq} } ,
\end{equation}
which coincides with the result obtain from the Boltzmann equation in
Eq.~\eqref{eq:Nhot} of the main text.

\subsection{El-ph self-energy and damping rate}

In order to obtain an expression for the inverse phonon lifetime due to el-ph
scattering, it is useful to express the imaginary part of the retarded
self-energy in terms of the greater and lesser self-energies as
\begin{equation}
  \tau_{\lambda\bq}^{-1} = -2 \text{Im}\, 
      \Pi_{\lambda\bq}^r \big\vert_{\omega=\omega_{\lambda\bq}}
  = i \left( \Pi_{\lambda\bq}^> - \Pi_{\lambda\bq}^< \right) 
  \big\vert_{\omega=\omega_{\lambda\bq}},
\end{equation}
where, as we shall see below, the two terms account for absorption and emission
processes, respectively.

To lowest order in the el-ph interaction, the phonon self-energy is given by the
\emph{bare} polarization operator times the square of the el-ph
interaction~\cite{Mahan}. In terms of the electronic Keldysh GFs, we can write
the self-energy as
\begin{equation}
   \Pi_{\lambda\bq}(\tau,\tau') = -i \abs{g_{\lambda\bq}}^2
       \sum_{\bk\sigma} G_{\bk + \bq}(\tau,\tau') G_{\bk}(\tau',\tau) ,
\end{equation}
where $X(\tau,\tau')$ denotes quantities with the time arguments ordered on the
Keldysh contour. Using the Langreth rules~\cite{Jauho} for the analytic
continuation onto the real-time axis and Fourier transforming to frequency
domain, the following expression for the greater/lesser self-energy is obtained,
\begin{equation}
   \Pi_{\lambda\bq}^{>/<}(\omega) = -i \abs{g_{\lambda\bq}}^2
       \sum_{\bk\sigma} \int \! \frac{d\varepsilon}{2\pi} \,
       G_{\bk + \bq}^{>/<}(\varepsilon + \omega) G_{\bk}^{</>}(\varepsilon) .
\end{equation}
Here, $G_\bk^{>/<}(\varepsilon) = \pm i {1 - f(\varepsilon) \choose
  f(\varepsilon)} A_\bk(\varepsilon)$ is the \emph{bare} electronic
greater/lesser GF and $A_\bk(\varepsilon)= 2\pi \delta(\varepsilon -
\varepsilon_\bk)$ is the electronic spectral function. Using the
$\delta$-function identity $\int d\varepsilon \, \delta(\varepsilon + \omega -
\varepsilon_{\bk+\bq}) \delta(\varepsilon - \varepsilon_\bk) =
\delta(\varepsilon_{\bk+\bq} - \varepsilon_\bk - \omega ) $, the inverse phonon
lifetime is found to be
\begin{align}
  \label{eq:tau_phonon}
  \tau_{\lambda\bq}^{-1} & = \frac{2\pi}{\hbar} \abs{g_{\lambda\bq}}^2
  \sum_{\bk\sigma}
    \bigg[ 
      f(\varepsilon_\bk) \left\{1 - f(\varepsilon_{\bk + \bq}) \right\}
    \bigg. \nonumber \\ 
    & \quad 
    \bigg. - 
      f(\varepsilon_{\bk + \bq}) \left\{1 - f(\varepsilon_\bk) \right\}
    \bigg]  \delta(\varepsilon_{\bk + \bq} - \varepsilon_\bk - \hbar\omega_{\lambda\bq} ) 
    \nonumber  \\
    & = \frac{2\pi}{\hbar} \abs{g_{\lambda\bq}}^2
  \sum_{\bk\sigma} \delta(\varepsilon_{\bk + \bq} - \varepsilon_\bk - \hbar\omega_{\lambda\bq} ) 
  \nonumber \\
  & \quad \times \big[ f(\varepsilon_\bk) - f(\varepsilon_\bk + \hbar\omega_{\lambda\bq}) \big],
\end{align}
where the identities in Ref.~\onlinecite{footnote1} have been applied to reach
the result in the last line. In the first equality, the two terms which originate
from the greater and lesser self-energies, respectively, are seen to describe
absorption and emission of phonons.


\begin{thebibliography}{73}
\bibitem[{foo()}]{footnote1}
\bibinfo{note}{Using the identities $$ f_\bk \left( 1 - f_{\bk + \bq} \right)
  N_\bq = f_{\bk + \bq} \left(1 - f_\bk \right) \left(1 + N_\bq \right) $$ and
  $$ f_{\bk + \bq} \left( 1 - f_\bk \right) = N_\bq \left( f_\bk - f_{\bk +
  \bq} \right) $$ where $f_{\bk + \bq} = f(\varepsilon_\bk + \hbar\omega_\bq)$
  and $N_\bq = N(\hbar\omega_\bq )$ is understood.}
\end{thebibliography}
\end{document}